\documentclass[twocolumn,epjc3]{svjour3}  
\smartqed  
\RequirePackage{graphicx}
\usepackage{array} 
\usepackage{booktabs} 
\usepackage{multirow}
\usepackage{textcomp} 
\usepackage{graphicx}
\usepackage{amsmath}  
\usepackage{amssymb} 
\usepackage{hyperref}
\usepackage[switch,modulo]{lineno}
\DeclareMathOperator{\sgn}{sgn}

\journalname{Eur. Phys. J. C}
\begin{document}

\title{Final results of the LOPES radio interferometer for cosmic-ray air showers}

\titlerunning{LOPES: Final Results}        

\author{{W.D.~Apel}\thanksref{addr_a}
        \and
        {J.C.~Arteaga-Vel\'azquez}\thanksref{addr_b}
        \and
        {L.~B\"ahren}\thanksref{addr_c}
        \and
        {K.~Bekk}\thanksref{addr_a}
        \and
        {M.~Bertaina}\thanksref{addr_d}
        \and
        {P.L.~Biermann}\thanksref{addr_e}
        \and
        {J.~Bl\"umer}\thanksref{addr_a, addr_f}
        \and
        {H.~Bozdog}\thanksref{addr_a}
        \and
        {E.~Cantoni}\thanksref{addr_d, addr_h}
        \and
        {A.~Chiavassa}\thanksref{addr_d}
        \and
        {K.~Daumiller}\thanksref{addr_a}
        \and
        {V.~de~Souza}\thanksref{addr_i}
        \and
        {F.~Di~Pierro}\thanksref{addr_d}
        \and
        {P.~Doll}\thanksref{addr_a}
        \and
        {R.~Engel}\thanksref{addr_a, addr_f}
        \and
        {H.~Falcke}\thanksref{addr_c, addr_e, addr_j}
        \and
        {B.~Fuchs}\thanksref{addr_f}
        \and
        {H.~Gemmeke}\thanksref{addr_l}
        \and
        {C.~Grupen}\thanksref{addr_m}
        \and
        {A.~Haungs}\thanksref{addr_a}
        \and
        {D.~Heck}\thanksref{addr_a}
        \and
        {J.R.~H\"orandel}\thanksref{addr_j}
        \and
        {A.~Horneffer}\thanksref{addr_e}
        \and
        {D.~Huber}\thanksref{addr_f}
        \and
        {T.~Huege}\thanksref{addr_a}
        \and
        {P.G.~Isar}\thanksref{addr_n}
        \and
        {K.-H.~Kampert}\thanksref{addr_k}
        \and
        {D.~Kang}\thanksref{addr_a}
        \and
        {O.~Kr\"omer}\thanksref{addr_l}
        \and
        {J.~Kuijpers}\thanksref{addr_j}
        \and
        {K.~Link}\thanksref[*, 1]{addr_a}
        \and
        {P.~{\L}uczak}\thanksref{addr_o}
        \and
        {M.~Ludwig}\thanksref{addr_f}
        \and
        {H.J.~Mathes}\thanksref{addr_a}
        \and
        {M.~Melissas}\thanksref{addr_f}
        \and
        {C.~Morello}\thanksref{addr_h}
        \and
        {S.~Nehls}\thanksref{addr_p}
        \and
        {J.~Oehlschl\"ager}\thanksref{addr_a}
        \and
        {N.~Palmieri}\thanksref{addr_f}
        \and
        {T.~Pierog}\thanksref{addr_a}
        \and
        {J.~Rautenberg}\thanksref{addr_k}
        \and
        {H.~Rebel}\thanksref{addr_a}
        \and
        {M.~Roth}\thanksref{addr_a}
        \and
        {C.~R\"uhle}\thanksref{addr_l}
        \and
        {A.~Saftoiu}\thanksref{addr_g}
        \and
        {H.~Schieler}\thanksref{addr_a}
        \and
        {A.~Schmidt}\thanksref{addr_l}
        \and
        {S.~Schoo}\thanksref{addr_a}
        \and
        {F.G.~Schr\"oder}\thanksref[*, 1, 17]{e1, addr_a, addr_q}
        \and
        {O.~Sima}\thanksref{addr_g, addr_r}
        \and
        {G.~Toma}\thanksref{addr_g, addr_s}
        \and
        {G.C.~Trinchero}\thanksref{addr_h}
        \and
        {A.~Weindl}\thanksref{addr_a}
        \and
        {J.~Wochele}\thanksref{addr_a}
        \and
        {J.~Zabierowski}\thanksref{addr_o}
        \and
        {J.A.~Zensus}\thanksref{addr_e} - LOPES Collaboration
}

\thankstext[*]{e1}{e-mail: frank.schroeder@kit.edu ; katrin.link@kit.edu}


\institute{{Institute for Astroparticle Physics (IAP) [formerly, Institute for Nuclear Physics (IKP)], Karlsruhe Institute of Technology (KIT), Karlsruhe, Germany}\label{addr_a}
           \and
           {Instituto de F\'isica y Matem\'aticas, Universidad Michoacana, Morelia, Michoac\'an, Mexico}\label{addr_b}
           \and
           {ASTRON, Dwingeloo, The Netherlands}\label{addr_c}
           \and
           {Dipartimento di Fisica, Universit\`a degli Studi di Torino, Torino, Italy}\label{addr_d}
           \and
           {Max-Planck-Institut f\"ur Radioastronomie, Bonn, Germany}\label{addr_e}
           \and
           {Institute for Experimental Particle Physics (ETP), Karlsruhe Institute of Technology (KIT), Karlsruhe, Germany}\label{addr_f}
           \and
           {National Institute of Physics and Nuclear Engineering, Bucharest-Magurele, Romania}\label{addr_g}
           \and
           {Osservatorio Astrofisico di Torino, INAF Torino, Torino, Italy}\label{addr_h}
           \and
           {Instituto de F\'isica de S\~ao Carlos, Universidade de S\~ao Paulo, S\~ao Carlos, Brasil}\label{addr_i}
           \and
           {Department of Astrophysics, Radboud University Nijmegen, AJ Nijmegen, The Netherlands}\label{addr_j}
           \and
           {Department of Physics, Bergische Universit\"at Wuppertal, Wuppertal, Germany}\label{addr_k}
           \and
           {Institut f\"ur Prozessdatenverarbeitung und Elektronik, Karlsruhe Institute of Technology (KIT), Karlsruhe, Germany}\label{addr_l}
           \and
           {Faculty of Natural Sciences and Engineering, Universit\"at Siegen, Siegen, Germany}\label{addr_m}
           \and
           {Institute of Space Science, Bucharest-Magurele, Romania}\label{addr_n}
           \and
           {Department of Astrophysics, National Centre for Nuclear Research, {\L}\'{o}d\'{z}, Poland}\label{addr_o}
           \and
           {Studsvik Scandpower GmbH, Hamburg, Germany}\label{addr_p}
           \and
           {Bartol Research Institute, Department of Physics and Astronomy, University of Delaware, Newark DE, USA}\label{addr_q}
           \and
           {Department of Physics, University of Bucharest, Bucharest, Romania}\label{addr_r}
         \and
           \emph{now at: European Commission, Joint Research Centre (JRC), Karlsruhe, Germany} \label{addr_s}
}

\date{preprint; accepted by EPJ C for publication}

\maketitle

\begin{abstract}
LOPES, the LOFAR prototype station, was an antenna array for cosmic-ray air showers operating from 2003 - 2013 within the KASCADE-Grande experiment. 
Meanwhile, the analysis is finished and the data of air-shower events measured by LOPES are available with open access in the KASCADE Cosmic Ray Data Center (KCDC).
This article intends to provide a summary of the achievements, results, and lessons learned from LOPES.
By digital, interferometric beamforming the detection of air showers became possible in the radio-loud environment of the Karlsruhe Institute of Technology (KIT). 
As a prototype experiment, LOPES tested several antenna types, array configurations and calibration techniques, and pioneered analysis methods for the reconstruction of the most important shower parameters, i.e., the arrival direction, the energy, and mass-dependent observables such as the position of the shower maximum. 
In addition to a review and update of previously published results, we also present new results based on end-to-end simulations including all known instrumental properties. 
For this, we applied the detector response to radio signals simulated with the CoREAS extension of CORSIKA, and analyzed them in the same way as measured data.
Thus, we were able to study the detector performance more accurately than before, including some previously inaccessible features such as the impact of noise on the interferometric cross-correlation beam. 
These results led to several improvements, which are documented in this paper and can provide useful input for the design of future cosmic-ray experiments based on the digital radio-detection technique.

\end{abstract}

\section{Introduction}
What is the potential of the digital radio detection technique for high-energy astroparticle physics?
To answer this question, LOPES (the LO{\small{FAR}} PrototypE Station) was deployed in the early 2000's and operated for about one decade. 
Meanwhile the experiment is concluded and data analysis finished. 
Therefore, it is time to summarize: What is the answer to the original question? What has been achieved? What else have we learned, what new questions emerged, and how is the field continuing?

Thanks to prototype experiments such as LOPES, digital radio detection has been established as an additional technique for the measurement of cosmic-ray air showers. 
First measurements of the radio emission were performed with analog detectors more than fifty years ago \cite{jelley,Allan1971,Vedeneev:2009zz}.
However, their accuracy for the cosmic-ray energy and the position of the shower maximum was limited by the analog technology used and by the incomplete theoretical understanding of the radio emission of air showers. 
Nevertheless, the community maintained a certain level of interest in the technique and continued experimental efforts regarding the radio detection of air showers \cite{HuegeReview2016,SchroederReview2016}.
A technical breakthrough was achieved in the 2000's with the digital antenna arrays LOPES \cite{FalckeNature2005} and CODALEMA \cite{ArdouinBelletoileCharrier2005} because the digitally saved data enabled sophisticated post-processing and computing-extensive analyses of the measurements. 
In parallel, progress was made on a variety of simulation and calculation tools \cite{HuegeCoREAS_ARENA2012,Alvarez_ZHAires_2012,SELFAS2_ARENA2012,Werner_EVA2012}, whose latest generations are generally able to reproduce measured radio signals.
Together with the second generation of digital antenna arrays such as AERA \cite{HuegeAERA_UHECR2018}, LOFAR \cite{SchellartLOFAR2013}, TREND \cite{TREND2019}, and Tunka-Rex \cite{TunkaRex_Xmax2016}, the measurements of LOPES and CODALEMA were critical to improve the understanding of the emission processes, in particular by comparing the measured radio signals with the predictions of Monte Carlo simulations of extensive air showers and to the data of the co-located particle detector arrays.  

Because LOPES was triggered by the co-located particle-detector array KASCADE-Grande, a comparison of radio and particle measurements of the same air showers demonstrated that radio detection provides an accuracy for the shower direction \cite{NiglDirection2008} and energy \cite{2014ApelLOPES_MassComposition} approximately equal to that of the particle-detector array.
As prototype station in the radio-loud environment of the Karlsruhe Institute of Technology, the precision of LOPES was, however, limited \cite{2014ApelLOPES_MassComposition}.
With improved methods and at more radio-quiet sites, the successor arrays LOFAR in the Netherlands \cite{BuitinkLOFAR_Xmax2014}, the Auger Engineering Radio Array (AERA) of the Pierre Auger Observatory in Argentina \cite{AERA_XmaxMethods_ARENA2016}, and Tunka-Rex in Siberia \cite{TunkaRexXmaxPRD2018} have meanwhile achieved an accuracy comparable to the leading optical techniques for air showers even for the position of the shower maximum.

In general, radio detection combines several advantages of the classical detection techniques for air showers. 
Like the air-fluorescence and air-Cherenkov light of air showers, also the radio emission depends on the relatively well-understood electromagnetic shower component, i.e., the radio signal has an intrinsic sensitivity to the energy of the electromagnetic component and the position of the shower maximum. 
The radio technique is not restricted to clear nights, but works under almost any weather conditions. 
LOPES showed that only thunderclouds directly over the antenna array have a significant impact on the radio signal \cite{2011ApelLOPES_Thunderstorm}, which lowers the available measurement time by only a few percent. 
Radio detectors share their advantage of availability around the clock with the classical technique of secondary-particle detection at the ground. 
Thus, antennas arrays are the ideal companion to particle detectors, in particular to muon detectors that yield complementary information to the electromagnetic shower component measured via the radio signal. 
Moreover, as it was done at LOPES, a particle detector array can provide a trigger to the radio antennas, which facilitates the discrimination of air-shower radio signal against background.

Digital radio experiments \cite{NellesLOFAR_measuredLDF2015,AugerAERAenergy2015,KostuninTheory2015} also revealed some complications of the radio technique: 
Due to the interplay of two different emission mechanisms, the radio footprint on ground has a two-dimensional, asymmetric shape. 
The two proven emission mechanisms are the geomagnetic deflection of the electrons and positrons \cite{KahnLerche1966},\cite{FalckeGorham2003}, which is the dominant effect for most air-shower geometries, and the weaker Askaryan effect, i.e., radio emission due to the variation of the net charge of the shower front during the shower development \cite{Askaryan1962}.
Several radio arrays in radio-quiet environments have experimentally confirmed the coexistence of these two emission mechanisms \cite{AugerAERApolarization2014,CODALEMAchargeExcess2015,SchellartLOFARpolarization2014}.
At LOPES, this complication was neglected against the typical measurement uncertainties, i.e., the LOPES measurements were interpreted as if the radio emission were purely geomagnetic. 
The resulting inaccuracy was small compared to other uncertainties, e.g., resulting from the high radio background at the site. 
Another complication of the radio technique regards self-triggering, which turned out to be possible, but difficult, and is unnecessary when radio arrays are operated as extension to particle-detector arrays. 
Finally, we learned that fully efficient detection for all arrival directions requires a relatively dense antenna spacing below $100\,$m, but sparse arrays with kilometer-wide spacing still qualify for the detection of inclined air showers. 
Despite these complications, the principle advantage of a precise all-day sensitivity to the electromagnetic shower component make the radio technique a suitable extension to existing particle-detector arrays, and potentially even for stand-alone neutrino detection.

Given these prospects of the radio technique, we have decided to provide this review of previous LOPES results starting with a summary of the main results, followed by a description of the LOPES setup and its various configurations. 
The focus is on the latest results using end-to-end simulations based on CoREAS, which are published in detail in the last PhD thesis related to LOPES \cite{LinkPhDThesis2016}, and have partly been shown on conferences \cite{LOPES_ICRC2015_Link,LOPES_ICRC2015_Schroeder,LOPES_ICRC2017}. 
We also include an overview of lessons learned, in the hope that some mistakes might be avoided by future radio experiments.

\section{Overview of LOPES results}
The following list provides a quick overview of important results obtained by LOPES. Some of them are reviewed in more detail in the later sections.

\begin{itemize}
\item LOPES was the first digital radio array which unambiguously proved that the detection technique is suitable to study air showers generated by high-energy cosmic rays \cite{FalckeNature2005}.
\item LOPES confirmed earlier results \cite{Allan1971} that the dominant emission mechanism is due to the Lorentz force. The amplitude of the radio signal is correlated with the geomagnetic angle \cite{FalckeNature2005}, and the ratio of the signal of differently aligned antennas is consistent with the polarization pattern expected by geomagnetic emission \cite{IsarARENA_2008},\cite{HuberPhDThesis2014}.
\item Several very inclined events were detected by LOPES \cite{PetrovicLOPESinclined2007}. The slope of the lateral distribution was observed to become flatter with increasing shower inclination \cite{2013ApelLOPESlateralComparison}, i.e. the more inclined the shower the larger the radio footprint. This is in line with later results by ANITA \cite{ANITA_CR_PRL_2010} and by the Auger Engineering Radio Array \cite{AERAinclined2018}.
\item Closeby thunderstorm clouds with high atmospheric electric fields can significantly change the radio emission of air showers and can cause much higher radio-signal amplitudes than during normal weather conditions \cite{BuitinkLOPESthunderstorms2007,2011ApelLOPES_Thunderstorm}. 
\item Digital interferometry, in particular cross-correlation beamforming, is an effective way of lowering the detection threshold and identifying air-shower pulses against background \cite{Apel:2006pv_LOPESdistantEvents}.
\item LOPES pioneered methods for the calibration of the absolute amplitude using an external reference source \cite{NehlsHakenjosArts2007},\cite{2015ApelLOPES_improvedCalibration}\footnote{Due to the better insight and update  of the absolute calibration, numerical values for amplitude measurements reported in publications prior to 2015 need to be corrected by a factor of $2.6\pm0.2$ as explained in Ref.~\cite{2015ApelLOPES_improvedCalibration}. Apart from this change of the amplitude scale, all results remain valid.}, and for the continuous calibration of the relative timing by a reference beacon emitting sine-wave signals \cite{SchroederTimeCalibration2010}. The nanosecond relative timing accuracy was essential for digital interferometry.
\item Thanks to its absolute calibration, LOPES was able to test the radio-signal amplitudes predicted by the REAS \cite{Ludwig:2010pf} and CoREAS \cite{HuegeCoREAS_ARENA2012} radio extensions of the CORSIKA Monte-Carlo simulation code on an absolute level. Applying the improved amplitude calibration showed that not REAS, but the newer CoREAS is fully compatible to LOPES within its measurement uncertainties \cite{2015ApelLOPES_improvedCalibration}.
\item The lateral distribution of the radio signal falls less steeply than indicated by the historic results from the analog era \cite{Allan1971}. A Gaussian lateral distribution function \cite{2014ApelLOPES_MassComposition} describes LOPES data best, but a simple exponential LDF turned out to be sufficient for many applications \cite{2010ApelLOPESlateral,2013ApelLOPESlateralComparison}. In the latter case, the average exponential decay constant observed by LOPES is approximately $R_0 = 180\,$m.  
\item The wavefront of the radio emission by air showers was found to be of hyperbolic shape \cite{2014ApelLOPES_wavefront}, which subsequently was confirmed by LOFAR with more precise measurements \cite{CorstanjeLOFAR_wavefront2014}.
\item The frequency spectrum observed with LOPES can be described by a power law or exponential function falling towards higher frequencies \cite{NiglFrequencySpectrum2008}.
\item LOPES provided an experimental proof that the radio signal depends on the longitudinal shower development \cite{2012ApelLOPES_MTD}, and was able to measure the depth of the shower maximum for individual events. Although the measurement uncertainties were too large for an interpretation in terms of mass composition, LOPES showed that the lateral distribution \cite{2014ApelLOPES_MassComposition} and the wavefront \cite{2014ApelLOPES_wavefront} can each be used to reconstruct the position of the shower maximum.
\item Despite the high radio background at the LOPES site, LOPES achieved a competitive accuracy of better than $20\,\%$ for the energy of the primary particle and of better than $0.5^\circ$ for the arrival direction (details in this article). 

\end{itemize}

\begin{figure}
  \centering
  \includegraphics[width=0.99\linewidth]{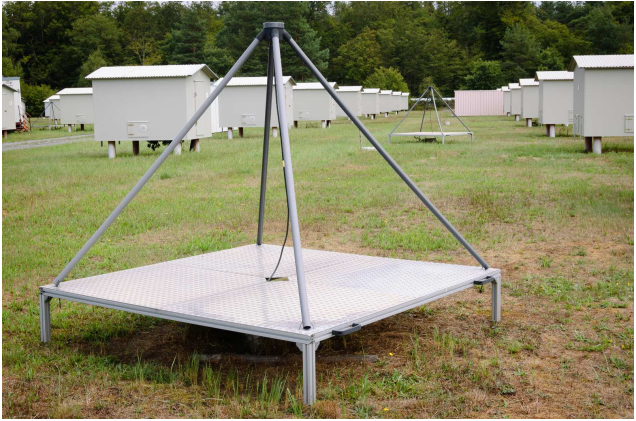}
  \caption{Photo of an inverted v-shape dipole antenna of the LOPES-30 setup in the KASCADE particle-detector array. 
}
  \label{fig_LOPESantenna}
\end{figure}

\begin{figure*}[t]
  \centering
  \includegraphics[width=0.99\linewidth]{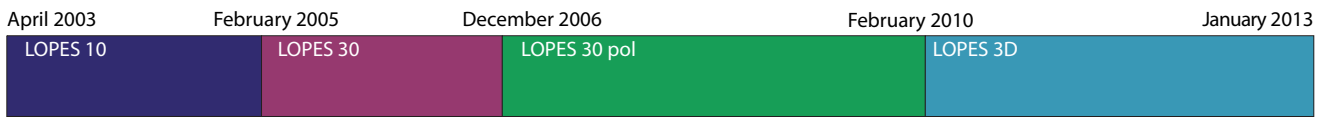}
  \caption{Timeline of the LOPES experiment: Starting with 10 east-west aligned antennas of the inverted v-shape dipole type (LOPES 10), LOPES was soon extended to 30 such antennas (LOPES 30). End of 2006, half of these antennas were rotated by $90^\circ$ to north-south alignment (LOPES 30 pol). Finally, all antennas were dismantled and 10 tripole antennas connected to the existing cables and DAQ infrastructure (LOPES 3D). 
  The results presented in this article are based on data acquired with the LOPES 30 and LOPES 30 pol setups.
  }
  \label{fig_timeline}
\end{figure*}

\begin{figure}
  \centering
  \includegraphics[width=0.99\linewidth]{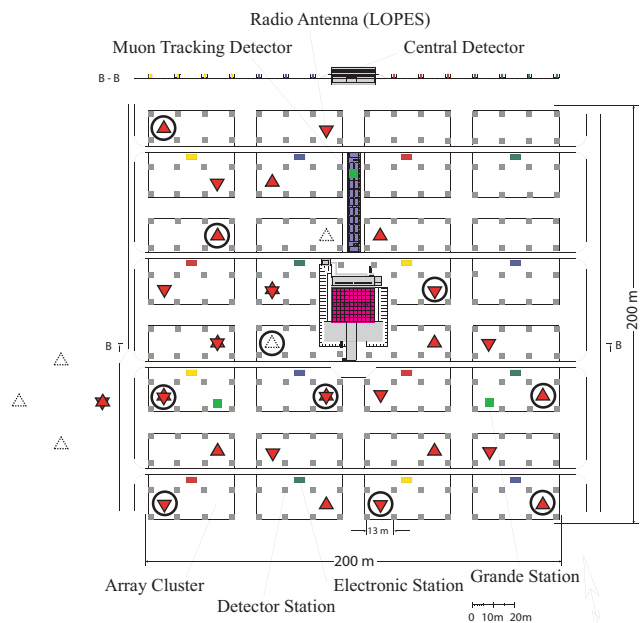}
  \caption{Map of the LOPES and KASCADE arrays: east-west aligned antennas are marked with upward triangles, north-south aligned antennas with downward triangles (i.e., stars are antenna stations equipped with both polarizations), circles mark the positions of the tripole stations deployed in 2009/2010, and dashed triangles mark those antennas dismantled in 2006 when deploying north-south aligned antennas.}
  \label{fig_LOPESexperiment}
\end{figure}

\section{The LOPES experiment}

This section describes the technical setup of the LOPES experiments and its data acquisition, the procedure for data analysis and end-to-end Monte Carlo simulations reproducing measured events. 
The section concludes with a description of the data set used for the results presented in this article and its public access via KCDC.

\subsection{Retrospective view on different experimental stages}
LOPES started in 2003 by deploying ten dipole-like LOFAR prototype antennas at the KASCADE array for cosmic-ray air showers (Fig.~\ref{fig_LOPESantenna}). 
Triggered by the particle-detector arrays KASCADE \cite{AntoniApelBadea2003} and KASCADE-Grande \cite{Apel2010KASCADEGrande}, LOPES soon detected the radio emission of air showers with energies above $10^{17}\,$eV \cite{FalckeNature2005}, and subsequently was extended to 30 antennas (Fig.~\ref{fig_timeline}). 
At the beginning, all LOPES antennas were aligned in east-west direction because, due to its dominant geomagnetic origin, the radio signal is on average strongest in the east-west direction. 

End of 2006, half of the then 30 antennas were rotated to align them in north-south direction.
The simultaneous measurements of the north-south and east-west aligned antennas provided additional evidence for the dominantly geomagnetic nature of the radio emission \cite{IsarARENA_2008}. 
Five antenna positions were equipped with both polarization directions, but due to restrictions in the cabling infrastructure the remaining ten east-west and ten north-south aligned antennas were placed at different positions (Fig.~\ref{fig_LOPESexperiment}). 
Since the radio signal and its polarization changes significantly on the scale of the antenna spacing ($26 - 37\,$m between adjacent antennas) \cite{ScholtenLOFAR_circPol2016}, this special separation of differently aligned antennas turned out to hamper a reconstruction of the signal polarization. 
At LOPES, we therefore decided to analyze the east-west and north-south aligned antennas separately. 
Because of the longer measurement time and the stronger signals, the event statistics was higher for the east-west aligned antennas \cite{2013ApelLOPESlateralComparison}, and most results are based on these east-west measurements -- including the new results presented in the following sections. 

In the winter of 2009/2010, the configuration of LOPES was changed again, i.e, the antennas were exchanged to ten 'tripole' stations, each consisting of three orthogonal dipole antennas aligned in east-west, north-south, and vertical direction \cite{2012ApelLOPES3D}. 
This setup was operating until 2013 when the whole KASCADE experiment was stopped and subsequently dismantled. 
Due to a new research facility close to the LOPES site, the background level increased substantially during the last phase of LOPES limiting the statistics of events above threshold. 
Nonetheless, statistical analyses and a few individual events with detectable signal in all three polarization directions again confirmed the general picture of the dominant geomagnetic emission \cite{HuberPhDThesis2014}. 
In addition to the overall increase of the background level, we noted that the background rises towards the horizon, as is expected for anthropogenic radio background.
Thus, no final conclusion was drawn whether such a more expensive setup with three polarization directions per station would have an advantage over a setup with two antennas per station.

\begin{figure*}[t]
  \centering
  \includegraphics[width=0.89\linewidth]{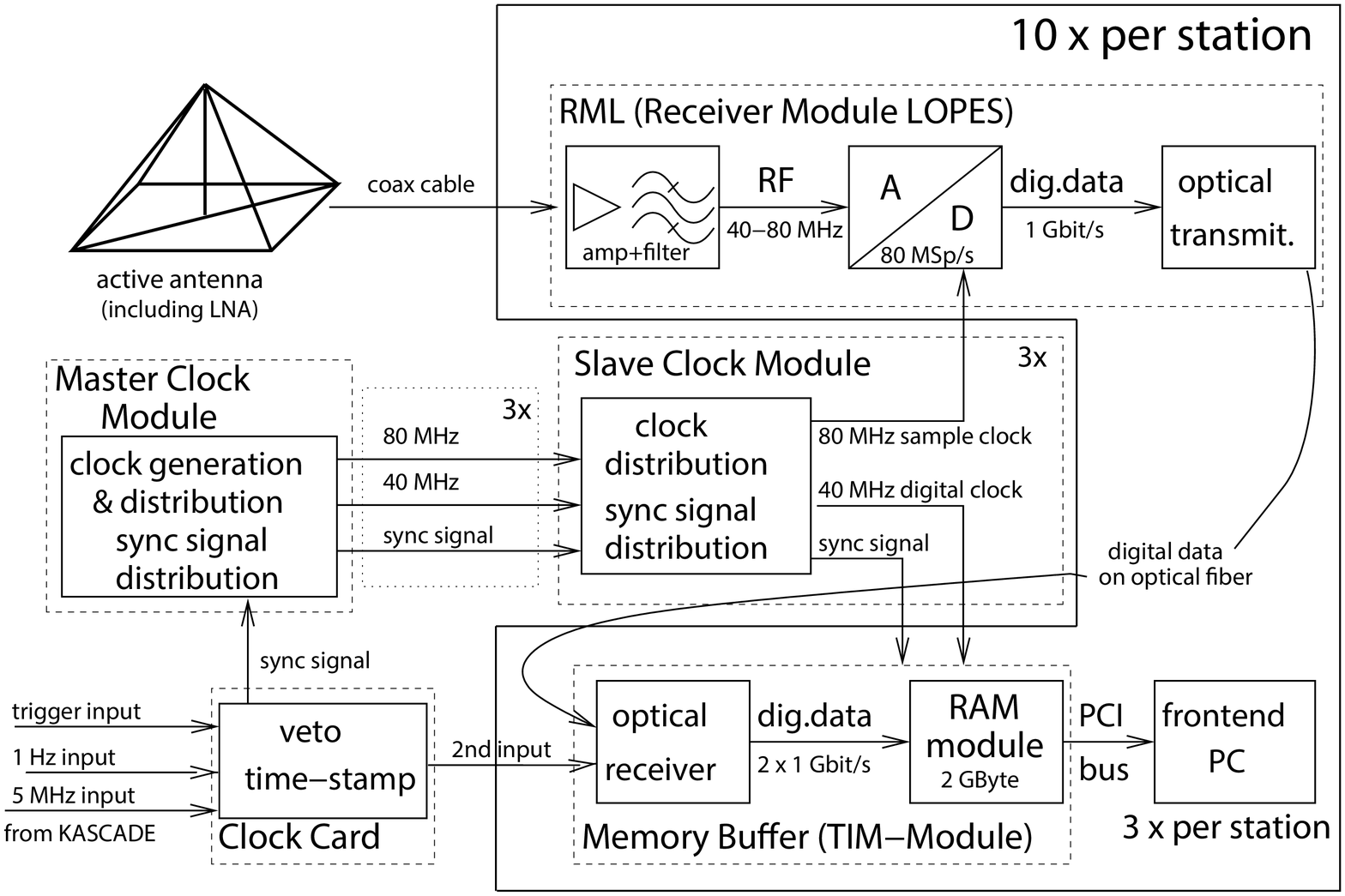}
  \caption{Data-acquisition system (DAQ) of LOPES organized in three stations of 10 channels, each. The signal of all antennas was digitized continuously in ring buffers. After receiving an external trigger, the signal of all 30 antennas were read out simultaneously and stored on one of nine local computers. Then, the data were combined to one event in a central computer and stored on a hard disk and as backup additionally on tape.
  }
  \label{fig_DAQ}
\end{figure*}


\subsection{Data-Acquisition System}
While the setup of the antennas was changed several times over the operation period of LOPES, the hardware of the data-acquisition remained the same. 
Technical details can be found, e.g., in reference \cite{2012ApelLOPES3D} and \cite{HornefferThesis2006}, and only the main features are described here.
 
The voltage measured at each antenna was continuously digitized and stored in a ring buffer, where the digitization of all antenna signals was synchronized by a common clock distributed via cables.    
Upon receiving a trigger from KASCADE or KASCADE-Grande, the buffers of all antennas were read out and combined to one event containing the coincident traces of all 30 antennas. 
Each event was stored on disk and combined during later analysis with the coincident KASCADE or KASCADE-Grande event providing the trigger, i.e, the radio signal measured by LOPES could be compared directly to the particle measurements of the same air shower. 

Due to limitations of analog-to-digital converters (ADCs) available at this time for a reasonable budget, a frequency band of a maximum width of $40\,$MHz had to be selected for LOPES.
The radio signal is strongest at frequencies below $100\,$MHz, because at these frequencies the wavelengths exceed the typical thickness of the particle front of air showers (today we know that the signal-to-noise ratio can be better at higher frequencies under some circumstances, and at the Cherenkov angle the radio emission extents even up to GHz frequencies as shown by the CROME experiment at the same site \cite{CROME_PRL2014}; see~\ref{sec_freqBand}).
Taking into account the knowledge when LOPES was built, the location of the frequency band of LOPES had been chosen to $40-80\,$MHz as a compromise between the Galactic background increasing towards lower frequencies and avoiding the FM band.

The radio signals received by the LOPES antennas were sampled in the second Nyquist domain with a nominal depth of 12 bits at a rate of $80\,$MHz (see Fig.~\ref{fig_DAQ}). 
Analog filters in the signal chain strongly suppressed the frequency range outside of the nominal band of $40-80\,$MHz.
This band was further reduced during analysis by digital filters to an effective band of $43-74\,$MHz to avoid systematic uncertainties because of slightly different cut-off frequencies of the individual filters. 
As the sampling conditions fulfill the Nyquist theorem, the radio signal between the samples in this band could later be retrieved by upsampling. 

Because the time synchronization of LOPES had glitches causing occasional offsets up to a few samples between antennas, we applied a dedicated method to improve the relative timing.
First, the carrier signals of TV transmitters in the measurement band were used and later a dedicated reference beacon emitting continuous sine waves \cite{SchroederTimeCalibration2010}. 
This external reference beacon ensured that the relative timing between different antennas was accurate to about $1\,$ns, which corresponds to a phase error of less than $30^\circ$ at $80\,$MHz. 
Together with an accurate measurement of the antenna positions by differential GPS this enabled the use of LOPES as a digital interferometer. 

The complete antenna array was calibrated by an external reference source several times per year starting in 2005 \cite{NehlsHakenjosArts2007}, achieving an absolute accuracy of the measured radio amplitude of about $16\,\%$ \cite{2015ApelLOPES_improvedCalibration}. 
Since the directional pattern of the antennas was not measured, the directional gain dependence of the antennas was taken from simulations which, however, came with some deficiencies (cf.~section \ref{sec_results}). 
Regarding the phase response of the LOPES signal chain, the largest effect was by the filters whose phase response was measured in the laboratory and was corrected for during analysis. 
However, we did not need to correct for the gain of the individual components, because LOPES features the end-to-end calibration of the absolute gain by the external reference source.

In summary, the data-acquisition system of LOPES was fully appropriate and fulfilled its purpose of providing reliable radio measurements of every triggered KASCADE(-Grande) event.

\begin{table}[t]
\centering
\caption{Properties of the LOPES experiment.}
\label{tab_LOPESproperties}
\begin{tabular}{ll}
\toprule
Location& $49^\circ 06'\,$N, $8^\circ 26'\,$E\\
Altitude& $110\,$m above sea level\\
Size& approx.~$0.04\,$km$^2$\\
Geomagnetic field~~~~& $B = 48.4\,$\textmu T, $\theta_B = 25.2^\circ$\\
Number of antennas~~~& up to $30$\\
Nominal band & $40-80\,$MHz \\ 
Effective band & $43-74\,$MHz \\ 
Trigger  & by co-located KASCADE(-Grande) \\ 
\bottomrule
\end{tabular}
\end{table}

\subsection{Analysis pipeline}
\label{sec_analysisPipeline}
Data analysis was performed by an open-source software 'CR-Tools' written in C++. 
The software as well as calibration and instrumental data of LOPES were made publicly accessible in a repository shared with LOFAR \cite{LOPESsoftware2017}. 
The LOPES software comes with different applications, e.g, for instrumental tests, calibration measurements, and a standard analysis pipeline for cosmic-ray air showers used for the results presented here.  
The individual steps of this pipeline are described in detail in various references \cite{LinkPhDThesis2016,SchroederThesis2011,HuegeARENA_LOPESSummary2010}. 

In the first step of the pipeline, the measurements were corrected for known instrumental properties such as the phase response of the filters, the simulated directional antenna pattern, and the total absolute gain obtained from end-to-end calibration measurements. 
In the second step of the pipeline, the data quality was enhanced by upsampling and by digitally removing narrow-band interferences: the frequency spectrum measured at each antenna was obtained by a fast Fourier transform (FFT) of the recorded traces.
Narrow-band lines in the frequency spectrum generally are of anthropogenic origin.
Consequently, they were suppressed.
This enhanced the signal-to-noise ratio and left the broad-band air-shower signal almost unchanged. 
Three of these suppressed lines in the frequency spectrum corresponded to the sine waves emitted by the dedicated reference beacon of LOPES \cite{SchroederTimeCalibration2010}, whose phasing was used to correct the relative timing between the individual antennas to an accuracy of about $1\,$ns. 

Upsampling was performed by the zero-padding method in the frequency domain: after a Fast Fourier Transform (FFT), zeroes were added to the frequency spectrum for the frequencies below the nominal band, i.e., for $0-40\,$MHz, for an upsampling factor of two, and also at higher frequencies until $n \times 40\,$MHz for an upsampling factor of $n > 2$.
Several cross-checks, e.g., with calibration pulses, had shown that high upsampling factors enabled a timing precision of better than $1\,$ns although the original samples were $12.5\,$ns apart. 
Generally, a higher upsampling increases the computation time for the analysis.
Therefore, for each analysis the upsampling factor was chosen such that the resulting sampling did not contribute significantly to the final uncertainties, which was given for most analyses at an upsampling factor of at least $n=8$, and for timing-sensitive analyses at an upsampling factor of at least $n=16$.

The interferometric method used at LOPES was cross-correlation (CC) beamforming, which was the next step in the pipeline: 
The traces of the individual antennas were shifted according to the arrival time of the radio wavefront.
This time shift depended on the arrival direction of the signal and on the shape of the radio wavefront. 
For the latter we used an hyperboloid centered around the shower axis with a variable angle $\rho$ between the shower plane and the asymptotic cone of the hyperboloid \cite{2014ApelLOPES_wavefront}. 
After the time shift of the individual traces, the CC beam $CC(t)$ was calculated as the sum of all pair-wise cross-correlations of shifted traces (details in Refs.~\cite{NiglFrequencySpectrum2008,Apel:2006pv_LOPESdistantEvents}):
\begin{equation}
CC(t) = \sgn(S(t))\sqrt{\frac{|S(t)|}{N_p}} ~~~ \mathrm{with}\\
S(t) = \sum_{i\ne j}^{N}s_i(t)s_j(t)
\end{equation}
with $N$ the number of antennas, $N_p$ the number of pairs (all combinations of two different antennas), and $s(t)$ the time-shifted signal in an individual antenna.

Because the CC beam features rapid oscillations, it was smoothed by block averaging over consecutive samples over $37.5\,$ns (i.e., $n \times 3$ samples for an upsampling factor of $n$).
A Gaussian was fitted to this smoothed CC-beam trace and the height of this Gaussian was used as measure for the amplitude of the CC beam. 
This smoothing made the reconstruction more robust, but lowered the amplitude of the CC beam.
Therefore, the CC-beam amplitudes depend on the reconstruction procedure and are  difficult to compare to other experiments.  

In an iterative fit procedure we searched for the maximum CC-beam amplitude by varying the cone angle $\rho$ of the wavefront and the arrival direction.\footnote{In an earlier version of the LOPES pipeline we used a spherical wavefront which corresponds to the approximating of a static point source. 
The maximization procedure of the CC-beam was similar, varying the arrival direction and the distance to the assumed point source (= radius of curvature) instead of the arrival direction and the cone angle. 
Such a method of searching for the point in the sky maximizing an interferometric beam was recently re-suggested in \cite{Schoorlemmer:2020low}. At LOPES, however, we switched to the hyberbolic wavefront model because it describes the measured and simulated events slightly better than a spherical wavefront model and also enabled a more precise $X_\mathrm{max}$ reconstruction \cite{2014ApelLOPES_wavefront}.}
To speed up the reconstruction procedure, we used the KASCADE(-Grande) reconstruction as initial value for the shower axis and search for the maximum in a range of $2.5^\circ$ around this initial value.
This range is more than five times larger than the direction accuracy of both arrays, KASCADE(-Grande) and LOPES, and we checked that the search range was large enough to avoid a bias due to the selection of the initial value. 
Hence, the maximization procedure yielded a reconstruction of the arrival direction as well as of the steepness of the radio wavefront. 

Using the arrival direction and cone angle found by maximizing the cross-correlation beam, we also form a power beam:
\begin{equation}
p(t) = \sqrt{\frac{1}{N} \sum_{i}^{N}s_i^2(t)}
\end{equation}
The fraction of the power and the CC-beam, the so-called excess beam, is a measure for the coherence of the signal \cite{Horneffer:2008hma}. 
Assuming that the air-shower pulse is mostly coherent in the individual antennas, incoherent contributions by background increase the value of the power beam, but not of the CC beam. 
Thus, the fraction of the total power contained in the CC beam is one of the quality criteria applied to the data set (cf.~Sec.~\ref{sec_dataset}).

Furthermore, by the time shift of the indiviudal traces that maximizes the CC beam, we knew the arrival time of the signal in each antenna. 
Thus, we could subsequently measure the signal amplitude at each individual antenna even very close to the noise level and without the need of applying additional quality cuts at the level of single antennas. 
Then, these amplitude measurements at the individual antennas were used for further analyses.

The amplitude measurements at the individual antennas are given in field strength per effective bandwidth, using an effective bandwidth of LOPES of $31\,$MHz (this means that values stated here need to be multiplied by $31\,$MHz to obtain the field strength in \textmu V/m in the effective band.) 
With some remaining limitations (see below), these amplitudes are easier to interpret than the CC beam and were used for the final step of the pipeline, which was the reconstruction of the lateral distribution. 
Although an exponential lateral distribution function (LDF) features an unphysical singularity at the shower axis, it turns out to provide a sufficient and simple empirical description of the LOPES measurements - given the significant uncertainties of typically $4-8\,$m in axis distance and at least $5\,\%$ in the individual amplitudes (much more at typical signal-to-noise ratios \cite{SchroederNoise2010}).
Hence, an exponential LDF with two free parameters was fitted to the amplitude, $\epsilon$, over distance to the shower axis, $d_\mathrm{axis}$:
\begin{equation}
 \epsilon(d_\mathrm{axis}) = \epsilon_{100} \exp (- \eta (d_\mathrm{axis} - 100\,\mathrm{m}))
 \label{eq_expLDF}
\end{equation}
where the amplitude at $100\,$m axis distance, $\epsilon_{100}$, is a good energy estimator, and the slope parameter $\eta$ is sensitive to the longitudinal shower development \cite{2012ApelLOPES_MTD}, as was predicted on the basis of simulations \cite{Huege:2008tn}. 

The lateral distribution was even better described by a Gaussian LDF which contains an additional free parameter.
We used the Gaussian LDF for the reconstruction of the energy and the position of the shower maximum \cite{2014ApelLOPES_MassComposition}. 
However, for the energy precision, the Gaussian LDF provided no significant improvement compared to the simpler exponential LDF \cite{LinkPhDThesis2016}.
We therefore use the simpler exponential LDF (Eq.~\ref{eq_expLDF}) for the analysis presented here.

\begin{figure*}[p]
  \centering
  \includegraphics[width=0.99\linewidth]{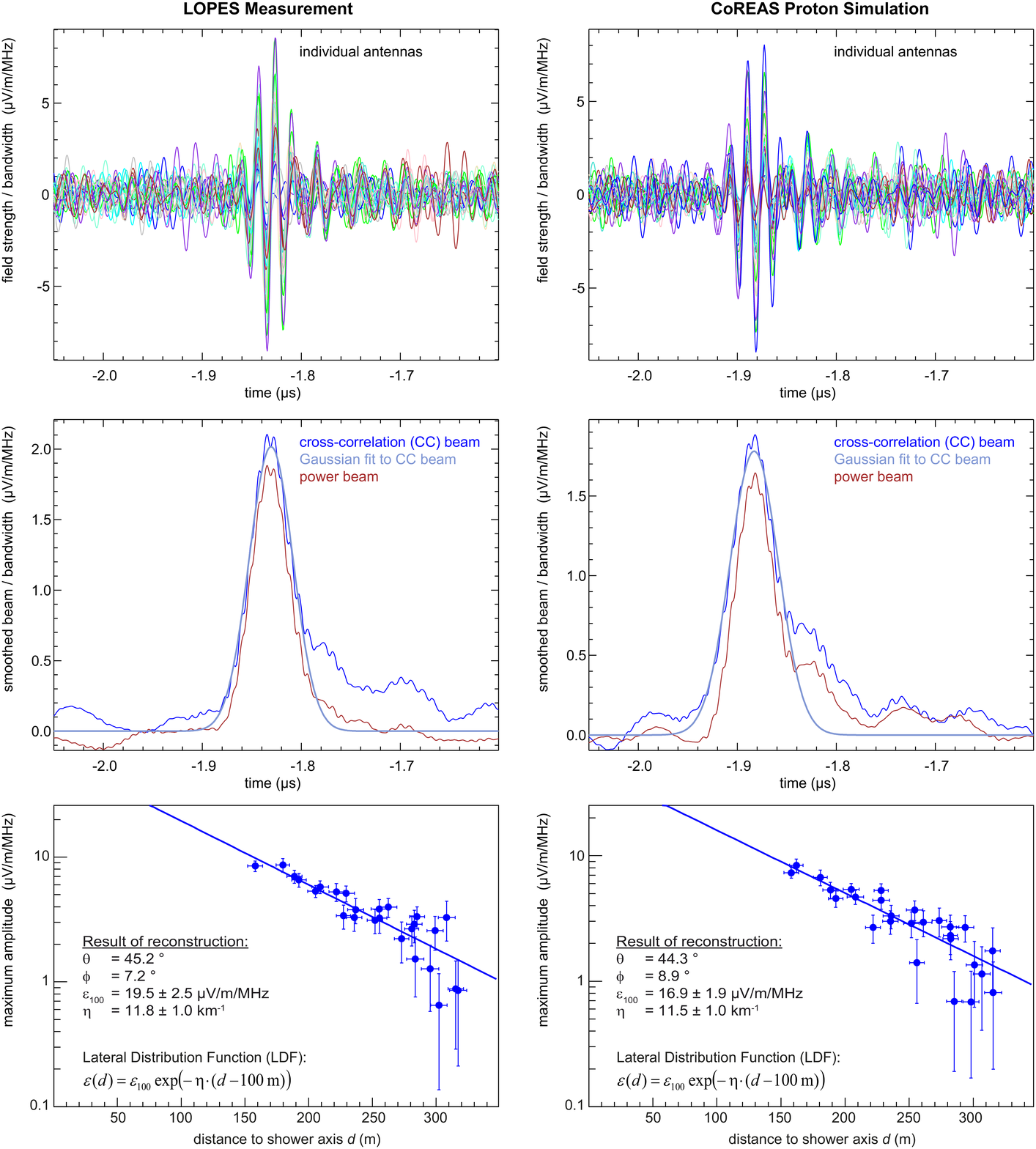}
  \caption{Example event measured by LOPES and simulated with CoREAS using the KASCADE-Grande reconstruction as input: energy $E = 3.4 \cdot 10^{17}\,$eV, azimuth $\phi = 8.6^\circ$, zenith $\theta = 44.1^\circ$. 
  This event is a best case example with high signal-to-noise ratio. Therefore, the signal is easily visible in all antennas.
  \emph{Left:} LOPES measurement. 
  \emph{Right:} CoREAS end-to-end simulation including the detector response. 
  \emph{Top:} Time series of the signal in the individual antennas (after the time shift for the hyperbolic beamforming in the shower direction). 
  \emph{Middle:} Cross-correlation beam and power beam after block-averaging over $37.5\,$ns for smoothing. 
  \emph{Bottom:} Lateral distribution of the maximum instantaneous amplitude (peak of Hilbert envelope) in each antenna and the fitted LDF (figure from Ref.~\cite{LOPES_ICRC2017}).}
  \label{fig_exampleEvent}
\end{figure*}

\begin{figure*}[t]
  \centering
  \includegraphics[width=0.49\linewidth]{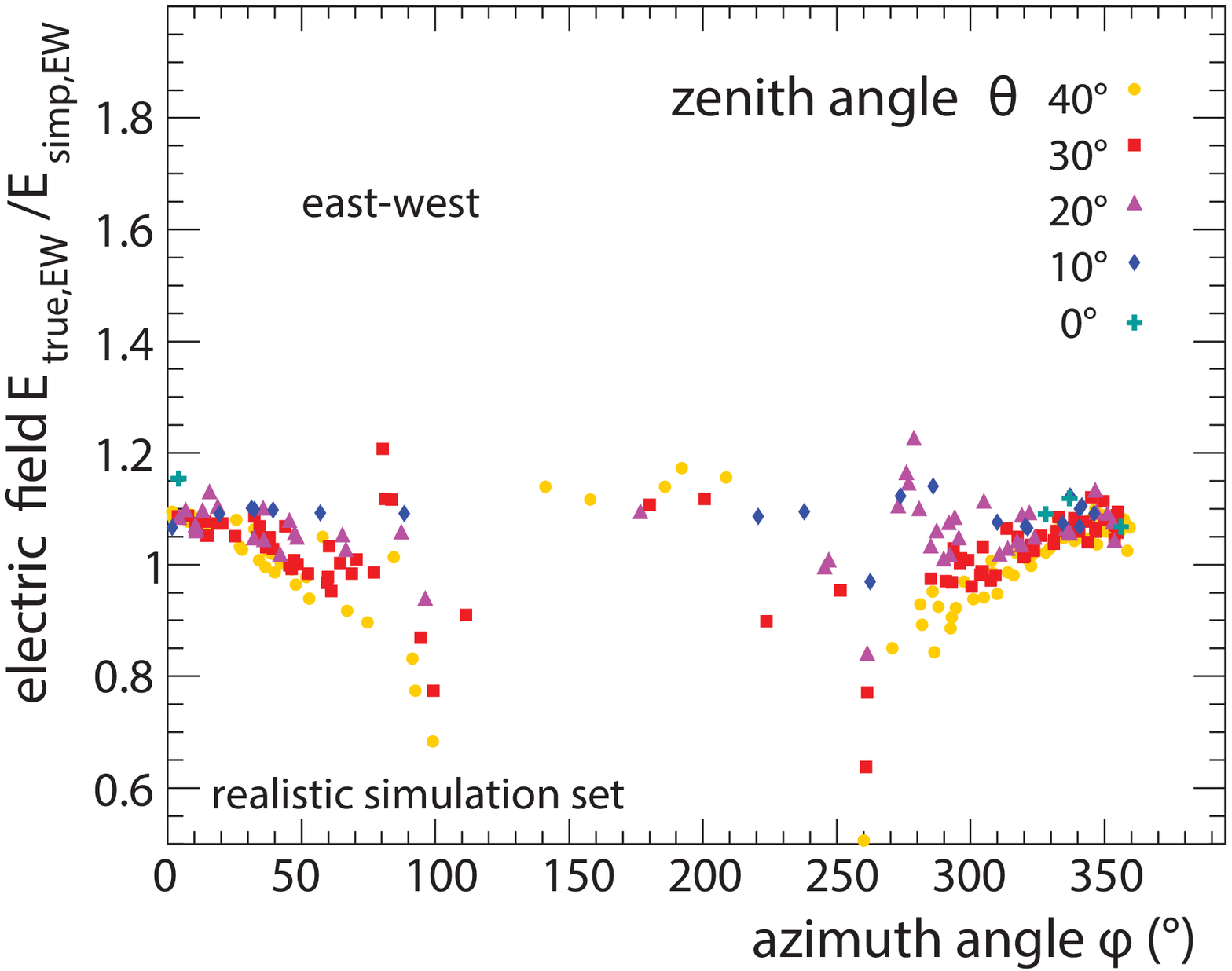}
  \hfill
  \includegraphics[width=0.49\linewidth]{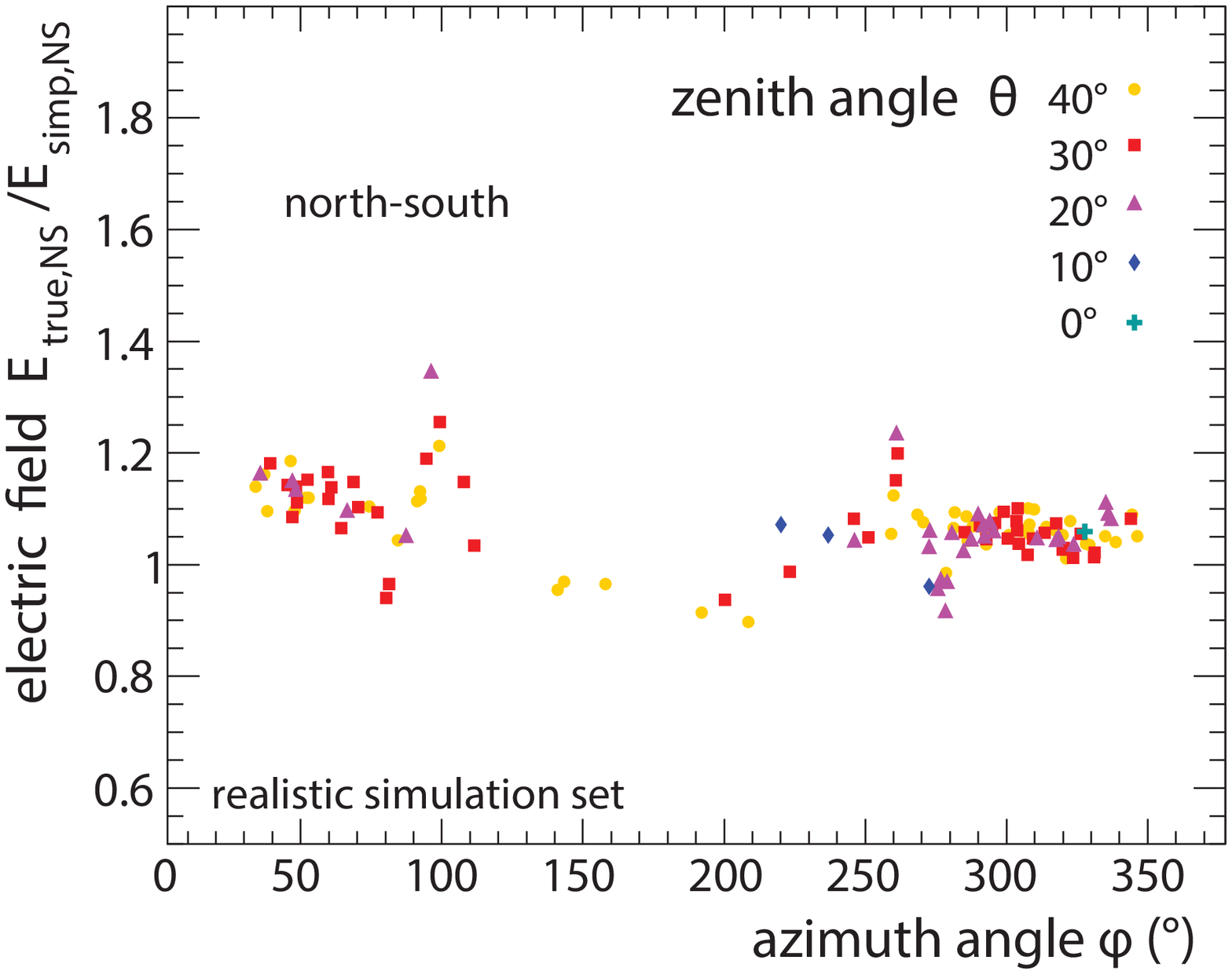}
  \caption{Ratio between the true polarization components in individual antennas and the values reconstructed by the standard analysis pipeline of LOPES for end-to-end CoREAS simulations including all known detector effects for the east-west (EW) and north-south (NS) aligned antennas. 
  While for individual events the deviation can be as large as a factor of 2, in most cases the reconstructed values are only slightly smaller than the true ones. 
  The average bias is only $(4.5 \pm 0.5)\,\%$ for the east-west component and $(7.4 \pm 0.6)\,\%$ for the north-south component.
  }
  \label{fig_reconstructionBias}
\end{figure*}

\begin{table*}[t]
\centering
\caption{Average bias due to noise determined by comparing the end-to-end simulations with and without noise for the east-west (EW) and north-south (NS) polarization components, respectively. 
The stated values were calculated as (1 - without noise / with noise) for values given in \% and as (without noise - with noise) for the slope parameter $\eta$. 
For the parameters of the lateral distribution, $\epsilon_{100}$ and $\eta$, also the biases due to the simplified reconstruction method of the electric field used by LOPES are stated: (1- true / reconstructed) and (true - reconstructed), respectively. 
Such a bias cannot be determined for parameters of the cross-correlation beamforming, since it implicitly includes the reconstruction simplifications and, thus, no 'true' values without bias are available from the CoREAS simulations.
The $\pm$ indicates the standard deviation, which for the bias due to noise can be interpreted as average statistical uncertainty of the corresponding quantity due to noise.}
\label{tab_noiseEffects}
\begin{tabular}{lcccc}
\toprule
& & bias due to reconstruction & bias due to noise & net bias \\ \midrule
\multirow{2}{*}{CC-beam amplitude} & EW & & $-(1.9 \pm 5.9)\,\%$ & \\
& NS & & $-(3.7 \pm 6.5)\,\%$ & \\ \midrule
\multirow{2}{*}{cone angle $\rho_{CC}$} & EW & & $+(0.2 \pm 12.0)\,\%$ & \\ 
& NS & & $~~(0.0 \pm 12.1)\,\%$ & \\ \midrule
\multirow{2}{*}{lateral amplitude $\epsilon_{100}$} & EW & $+(1.5 \pm 8.1)\,\%$ & $-(1.1 \pm 7.3)\,\%$ & $+(1.9 \pm 10.6)\,\%$ \\ 
& NS & $+(6.6 \pm 7.3)\,\%$ & $-(1.9 \pm 7.3)\,\%$ & $+(3.7 \pm 14.4)\,\%$ \\ \midrule
\multirow{2}{*}{lateral slope $\eta$} & EW & $-(1.05 \pm 0.64)\,$km$^{-1}$ & $+(1.01 \pm 1.80)\,$km$^{-1}$ & $+(0.08 \pm 1.72)\,$km$^{-1}$\\
& NS & $-(1.02 \pm 0.85)\,$km$^{-1}$ & $+(0.52 \pm 1.70)\,$km$^{-1}$ &  $+(0.68 \pm 2.35)\,$km$^{-1}$\\ \bottomrule
\end{tabular}
\end{table*}

\subsection{End-to-end simulations}
\label{sec_reconstructionBias}
The latest feature implemented in the analysis software is the treatment of air-shower simulations in the same way as measured data. 
The radio signal of air-showers was calculated by the CoREAS extension of the CORSIKA Monte Carlo code \cite{HuegeCoREAS_ARENA2012}. 
Afterwards, all known instrumental effects were applied on simulated electric-field vectors at each antenna position, in particular the gain pattern of the antennas, the amplitude and phase characteristics of the signal chain, and the quantization of the signal implied by the resolution of the 12-bit ADC. 
The simulated signals were stored as traces with the LOPES sampling rate of $80\,$MHz in the same data format as real events (see Fig.~\ref{fig_exampleEvent} for a typical example event). 
Subsequently, the simulations were analyzed using the same standard analysis pipeline as for the measurements. 

Optionally measured noise was added to the simulated events. 
For this purpose, we used real background measured by the LOPES antennas. 
Thus, the performance of the LOPES experiment could be assessed using these end-to-end simulations. 
In particular the cross-correlation beamforming was studied with the simulations, and the measurements of the hard to interpret CC beam were compared quantitatively to the predictions of the CoREAS simulations. 

With the new end-to-end simulations, we were able to check the effect of a simplification made when comparing REAS and CoREAS simulations to LOPES measurements in earlier publications \cite{2013ApelLOPESlateralComparison}: 
when processing the simulations, we simply filtered the east-west and north-south polarization components of the simulated electric field to the effective band of LOPES, but ignored that LOPES was unable to measure these electric field components directly.
Due to the inverted v-shape of the LOPES antennas the east-west and north-south aligned antennas are also partially sensitive to vertically polarized signals.
In contrast to other experiments featuring two orthogonally aligned antennas at each station, the three components of the electric field vector could not be reconstructed at LOPES.
Since at most antenna positions only one antenna was available, e.g., east-west aligned, necessarily a simplifying assumption had to be made in the reconstruction of the radio signal.
Therefore, we used a simplified treatment of the deconvolution of the direction-dependent antenna pattern, which is described in detail in reference \cite{LinkPhDThesis2016}.
Nevertheless, with the new end-to-end simulations we treated simulations and measurements in the same way and were able to fully compare them with each other.

\begin{figure}
  \centering
  \includegraphics[width=0.99\linewidth]{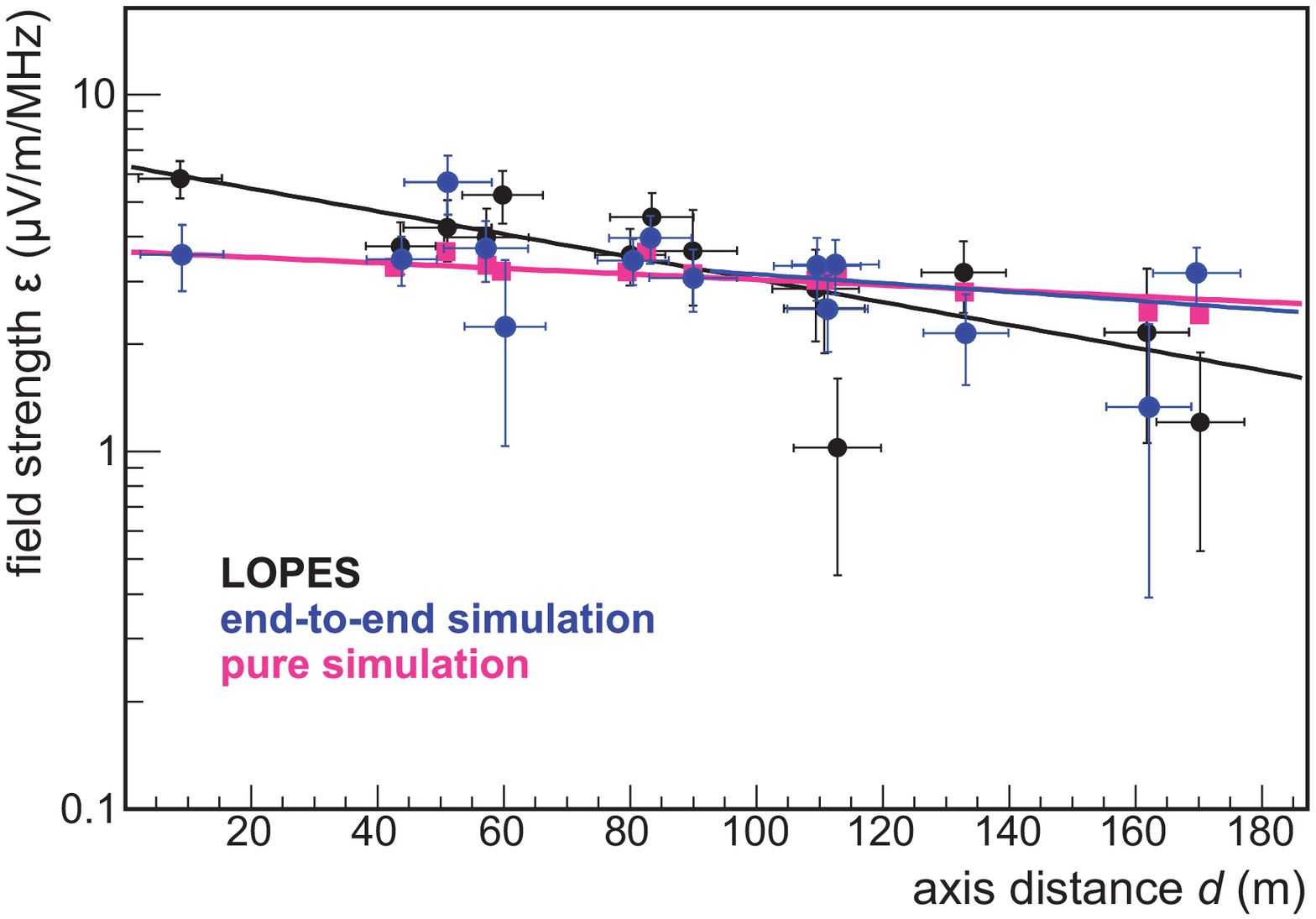}
  \caption{Lateral distribution of a typical example event measured by LOPES and simulated with CoREAS. 
  The application of the detector response (= end-to-end) impacts the CoREAS simulations only marginally. 
  Generally, a different slope for the simulated and measured lateral distribution is expected for individual events because the position of the shower maximum is subject to shower-to-shower fluctuation.
  }
  \label{fig_lateralExample}
\end{figure}

Furthermore, with the end-to-end simulations we were able to study the error made by the simplification. 
Since both, the polarization of the radio signal emitted by an air shower as well as the antenna gain, depend on the arrival direction, the size of the error is strongly arrival-direction dependent.
While for individual events the error can be as large as a factor of 2, on average the reconstructed values of the field strengths are only few percent lower than the true values of the simulated air showers.  
Figure \ref{fig_reconstructionBias} shows the dependence of this bias on the azimuth and zenith angle: each point corresponds to the arrival direction of a real air shower measured by LOPES (the distribution is nonuniform because the amplitude of the radio signal and the detection threshold of LOPES depended strongly on the arrival direction of the air shower relative to the geomagnetic field). 
For the large majority of events, the error by the simplification is smaller than $10\,\%$ and well within systematic uncertainties quoted in earlier publications.

In addition to the bias for the reconstructed field strength in individual antennas, we also studied the average effect of the simplified treatment of the antenna gain on the reconstructed lateral distribution (see Fig.~\ref{fig_lateralExample} for an example). 
By comparing the true values of the simulations with the result of the end-to-end simulations with and without noise, we discovered that noise introduces an additional bias on the amplitude and slope parameters, $\epsilon_{100}$ and $\eta$, respectively.
This bias is on top of a bias due to noise in individual antennas, which we had already corrected for in our standard analysis \cite{SchroederNoise2010}. 
For the amplitude parameter $\epsilon_{100}$, the size of each effect (noise bias and antenna-gain-simplification bias) is small relative to the dominating $16\,\%$ scale uncertainty of the amplitude calibration.
For the slope parameter $\eta$, the size of the individual biases are comparable to the measurement uncertainties, but the biases by noise and by the simplified treatment of the antenna gain partly compensate each other.
Overall, the mean net biases are small, but there is a relatively large spread, which reflects an event-by-event systematic uncertainty (see Table~\ref{tab_noiseEffects}). 
This implies that the measurements of individual events have to be interpreted with care while average values over dozens to hundreds of events should be affected only marginally.
Consequently, LOPES results published prior to this paper can be considered reliable. 

\begin{figure}
  \centering
  \includegraphics[width=0.99\linewidth]{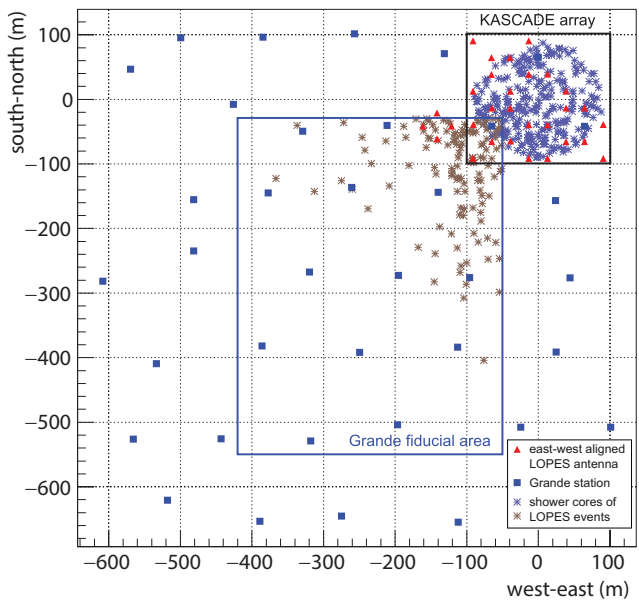}
  \caption{Impact point of the shower axis (core) of the events passing the quality cuts described in the text. 
  In addition to the KASCADE array (Fig.~\ref{fig_LOPESexperiment}) also the detector locations of the larger KASCADE-Grande array are shown.
  }
  \label{fig_eventMap}
\end{figure}

\begin{table}
\centering
\caption{Statistics of LOPES events used in this paper. The two subsets of LOPES events triggered and well reconstructed by KASCADE and KASCADE-Grande, respectively, overlap by a few events which is why the total number of events is less than the sum. 
For those measured events remaining after all quality cuts, also the statistics of corresponding showers simulated by CoREAS are shown that pass all quality cuts for both cases of a proton and iron nucleus as primary particle.}
\label{tab_LOPESstatistics}
\begin{tabular}{lccc}
\toprule
Cumulative Quality Cuts& KASCADE & Grande & Total\\
\midrule
$E>10^{17}\,$eV & 951 & 3042 & 3974\\
signal-to-noise ratio & 415 & 310 & 715\\
power in CC-beam & 345 & 245 & 582 \\ 
exclude thunderstorms & 339 & 239 & \textbf{570} \\ 
\midrule
simulated events & 302 & 162 & 464 \\ 
sim.~events with noise & 258 & 122 & \textbf{380} \\ 
\bottomrule
\end{tabular}
\end{table}


\subsection{Data set}
\label{sec_dataset}
\begin{figure}
  \centering
  \includegraphics[width=0.99\linewidth]{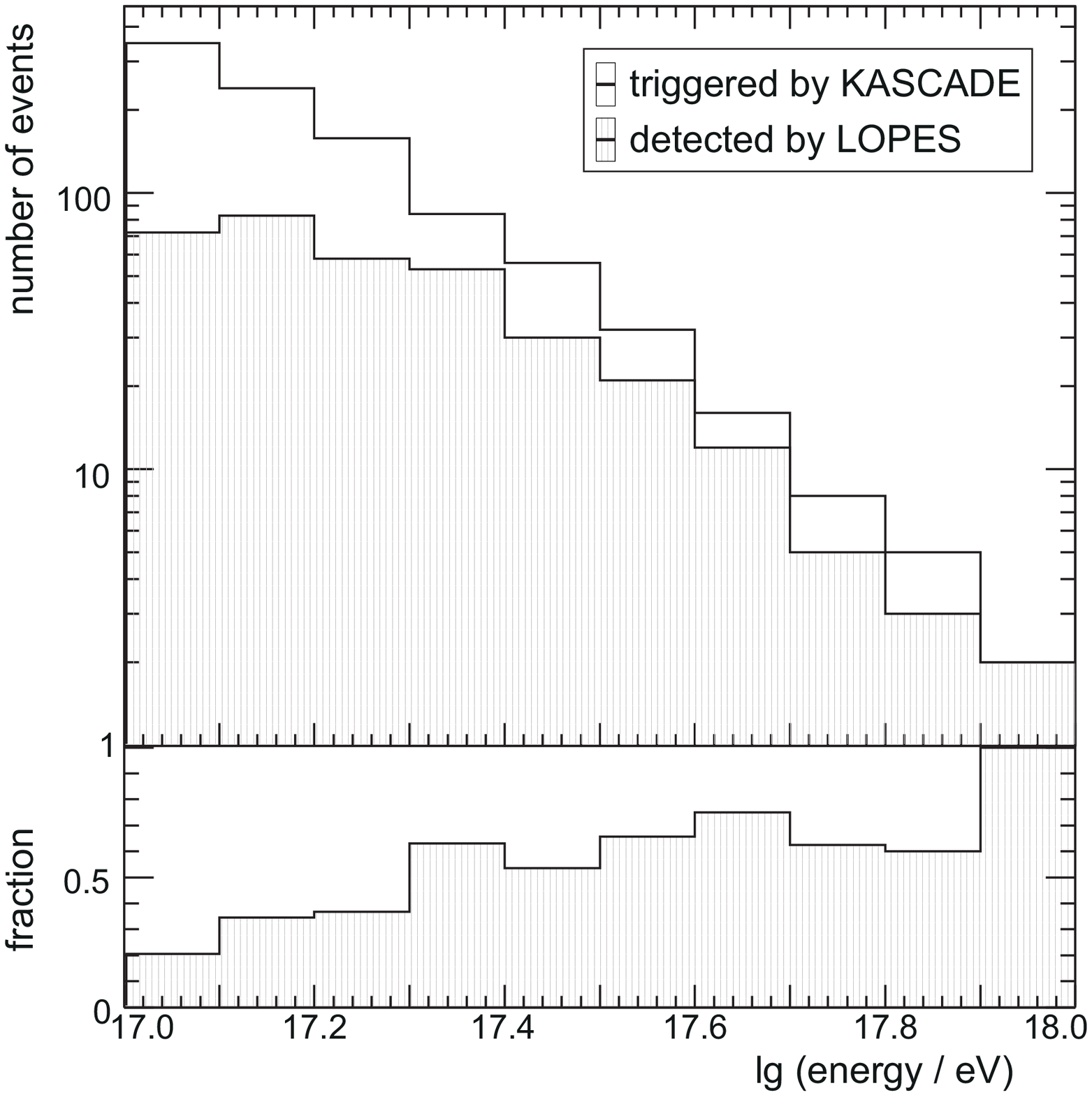}
  \caption{Top: Number of events triggered by the KASCADE array with a reconstructed energy above $10^{17}\,$eV (no event had an energy above $10^{18}\,$eV); and the number of events in the subset passing all LOPES quality cuts.
  Bottom: Fraction of the two event numbers which is a measure for the efficiency of LOPES.}
  \label{fig_efficiency}
\end{figure}

The LOPES data set consists of air-shower events triggered by the KASCADE array and its KASCADE-Grande extension. 
Both arrays provided a trigger to LOPES for all events with estimated energies $\gtrsim 10^{16.5}\,$eV, which was significantly lower than the detection threshold of LOPES around $10^{17}\,$eV.
We removed those events from the analysis which had a zenith angle larger than $45^\circ$ or which had their shower cores outside of the fiducial areas of the KASCADE or KASCADE-Grande array, respectively.
After applying these cuts, both arrays were fully efficient for all types of primary cosmic rays well below the relevant energy range, and we can safely assume that all events that had a radio signal passing the LOPES reconstruction were triggered. 
However, LOPES itself was not fully efficient, i.e., only a fraction of the triggered events passed the LOPES reconstruction (see Table~\ref{tab_LOPESstatistics} and Fig.~\ref{fig_efficiency})\footnote{Because the biases related to the partial efficiency of LOPES are difficult to estimate, we refrain from determining the absolute flux, energy spectrum, or mass composition.}.
Corresponding to the two particle-detector arrays providing the trigger, there are two data sets of LOPES events, KASCADE and Grande events, which have only little overlap (Fig.~\ref{fig_eventMap}).
Depending on the analysis, we either use both data sets combined, or only the KASCADE data set because those events have their core contained inside of the LOPES antenna array which allows for a higher quality of the event reconstruction.

For the present analysis we used data recorded by the east-west aligned v-shape dipole antennas from December 2005 to October 2009, because starting December 2005 LOPES featured an absolute amplitude calibration \cite{NehlsHakenjosArts2007}.
During this time almost 4000 well-reconstructed KASCADE and KASCADE-Grande events with an energy of at least $10^{17}\,$eV triggered LOPES and were propagated through the LOPES analysis pipeline.
To those events which passed the analysis pipeline without error, which implies, e.g., that the reconstruction of the arrival direction converged, we applied further quality cuts:
\begin{itemize}
\item The signal-to-noise ratio of the CC-beam must be greater than $14 \cdot \sqrt{N_\mathrm{ant}/30}$, with $N_\mathrm{ant}$ the number of antennas contributing to the measurement of the event. 
\item The CC beam must contain at least $80\,\%$ of the total power, which excluded events contaminated by background. 
\item To reject thunderstorm events, events with an atmospheric electric field of at least $3\,$kV/m were excluded. This cut was applied to events recorded after the installation of a local electric field mill on 24 August 2006.
\end{itemize}
After the quality cuts, 570 measured LOPES events remain in the analysis.

Using the KASCADE dataset, we were able to estimate the efficiency because the shower cores of these events were contained or very close to the LOPES antenna array (Fig.~\ref{fig_efficiency}).
For energies above $2\cdot10^{17}\,$eV, more than half of the LOPES events passed all of the quality cuts mentioned above.
For most of the Grande events, the shower cores were too distant from the LOPES array for a detectable radio signal.
More detailed discussions on the dependencies of the amplitude of the radio signal, e.g,. on the energy, the geomagnetic angle, and the distance to the shower axis can be found in many of the references cited in the introduction.

We also produced a library of CoREAS simulations using the energy, arrival direction, and shower core reconstructed by KASCADE(-Grande)\footnote{'KASCADE(-Grande)' refers to both, the KASCADE and the KASCADE-Grande data sets, at the same time.} as input. 
Using CORSIKA 7.3 with the hadronic interaction model QGSJet II.03, two simulations were created per LOPES event, one with a proton and one with an iron nucleus as primary particle.
Because different LOPES analyses used different selection criteria, versions of the analysis pipeline, and subsequent quality cuts, the exact data sets varied slightly over time and LOPES publication, and some showers were not included in the simulation library. 
Still, there is significant overlap between all selections, and for about $90\,\%$ of the measured events used here there are corresponding CoREAS simulations. 
Each simulation was processed twice through the standard analysis pipeline, once the pure simulated radio traces and once the simulated radio traces after adding randomly selected noise samples measured by LOPES.

After the analysis pipeline, the simulated events were subject to the same quality cuts as the measured events.
Since many of the measured events are close to the detection threshold, only a part of the corresponding simulated showers passed the quality cuts (the vice-versa situation does not happen because those showers missing the cuts for the measurements, were simply not simulated). 
In case of pure simulations without adding noise, 464 events passed all quality cuts for both, proton and iron, as primary particle (after removing 3 simulated events for which the fit of the lateral distribution failed).
In case of the simulations with measured noise added, 380 events passed all quality cuts for both, proton and iron, as primary particle.
These common events with both measured and simulated results available are the data set used for the results shown here.

The LOPES events were made available to the public in the \textit{KASCADE Cosmic Ray Data Center (KCDC)} \cite{KCDC2018} in November 2019 as part of the 'Oceanus' release.\footnote{Due to different versions of the KASCADE reconstruction software 'KRETA', the values available in KCDC may differ slightly from the ones used in the analysis presented in this paper.}. 
In addition to the reconstructed parameters of these LOPES events (up to 20 parameters per event and 4 parameters per antenna in an event), also the corresponding KASCADE-Grande data can be downloaded from KCDC as detailed in the user manual available on the KCDC website. 
The LOPES data can be found as subset of the KASCADE data in the KCDC data shop: \url{ https://kcdc.ikp.kit.edu/}

\section{Results}
\label{sec_results}

Using the end-to-end simulations, we first checked for the consistency of measurements and simulations. 
Generally, the CoREAS simulations describe the measured events well. 
Therefore, in a second step we used the simulations to study several features of our analysis procedure, in particular the relation between the amplitude at a specific lateral distance and the CC-beam amplitude.
Finally, we provide an update on earlier results regarding the direction and energy accuracy of LOPES and the sensitivity to the shower maximum.

\begin{figure*}[t]
  \centering
  \includegraphics[width=0.49\linewidth]{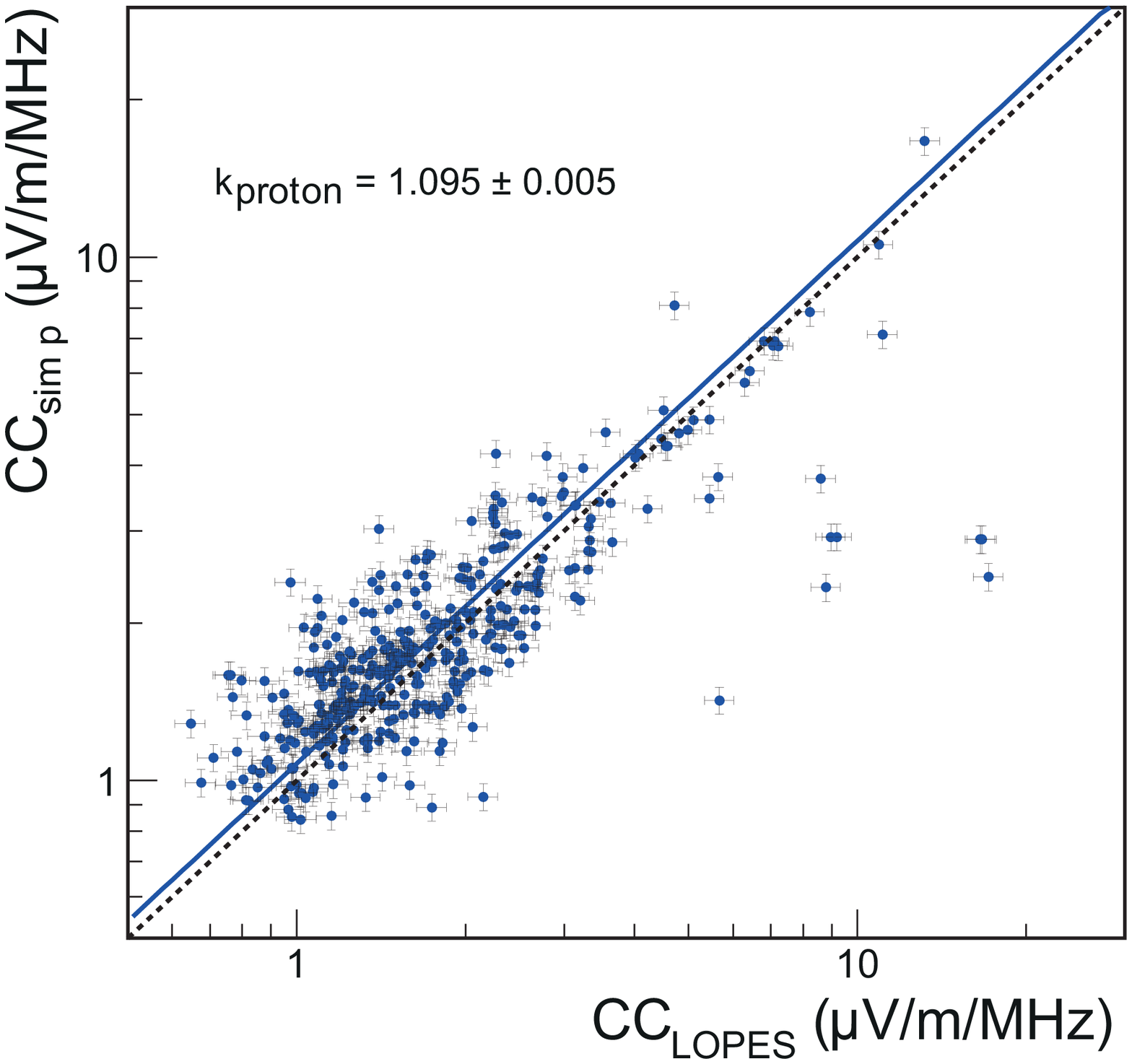}
  \hfill
  \includegraphics[width=0.49\linewidth]{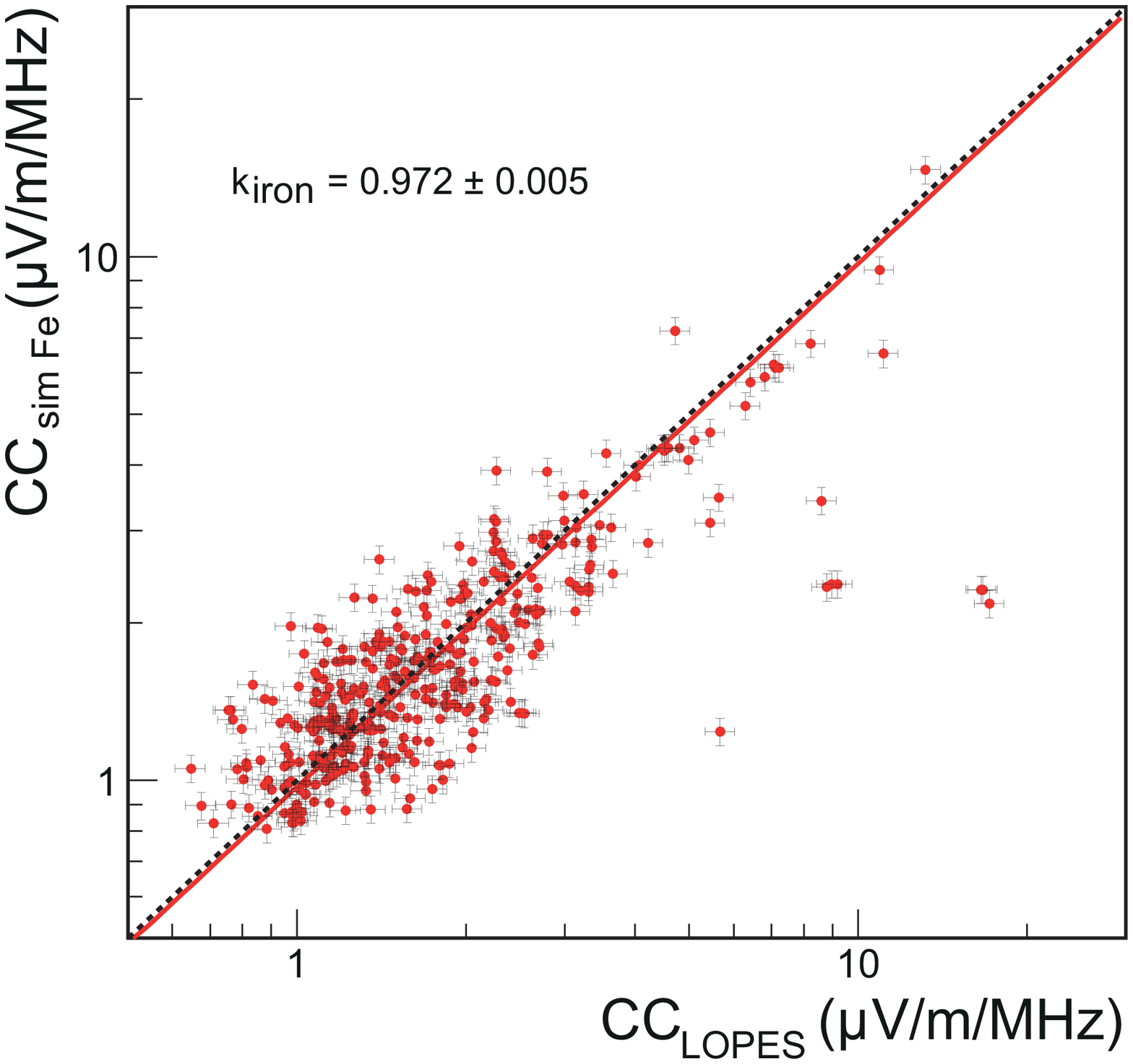}
  \caption{Event-by-event comparison of the cross-correlation-beam amplitude of the LOPES measurements and the end-to-end CoREAS proton and iron simulations (including noise; for east-west antennas). 
  The $k$ values are the offsets of the fitted dashed lines to the one-to-one correlation (solid lines), which is in both cases smaller than the calibration scale uncertainty of $16\,\%$. 
  The outliers are mostly the same events as in Fig.~\ref{fig_zenithDependence}.
  }
  \label{fig_CCbeamComparison}
\end{figure*}

\begin{figure*}[t]
  \centering
  \includegraphics[width=0.49\linewidth]{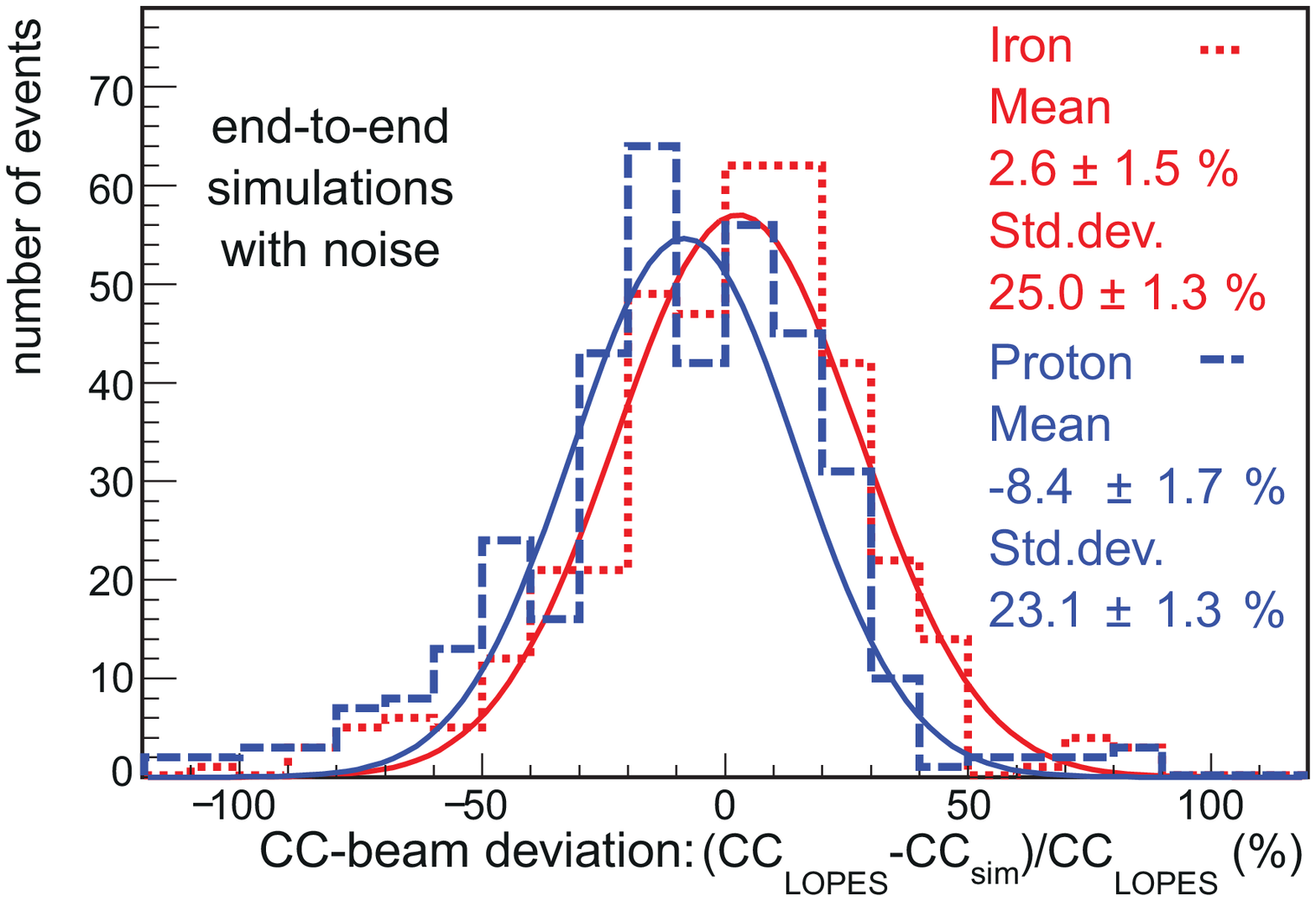}
  \hfill
  \includegraphics[width=0.49\linewidth]{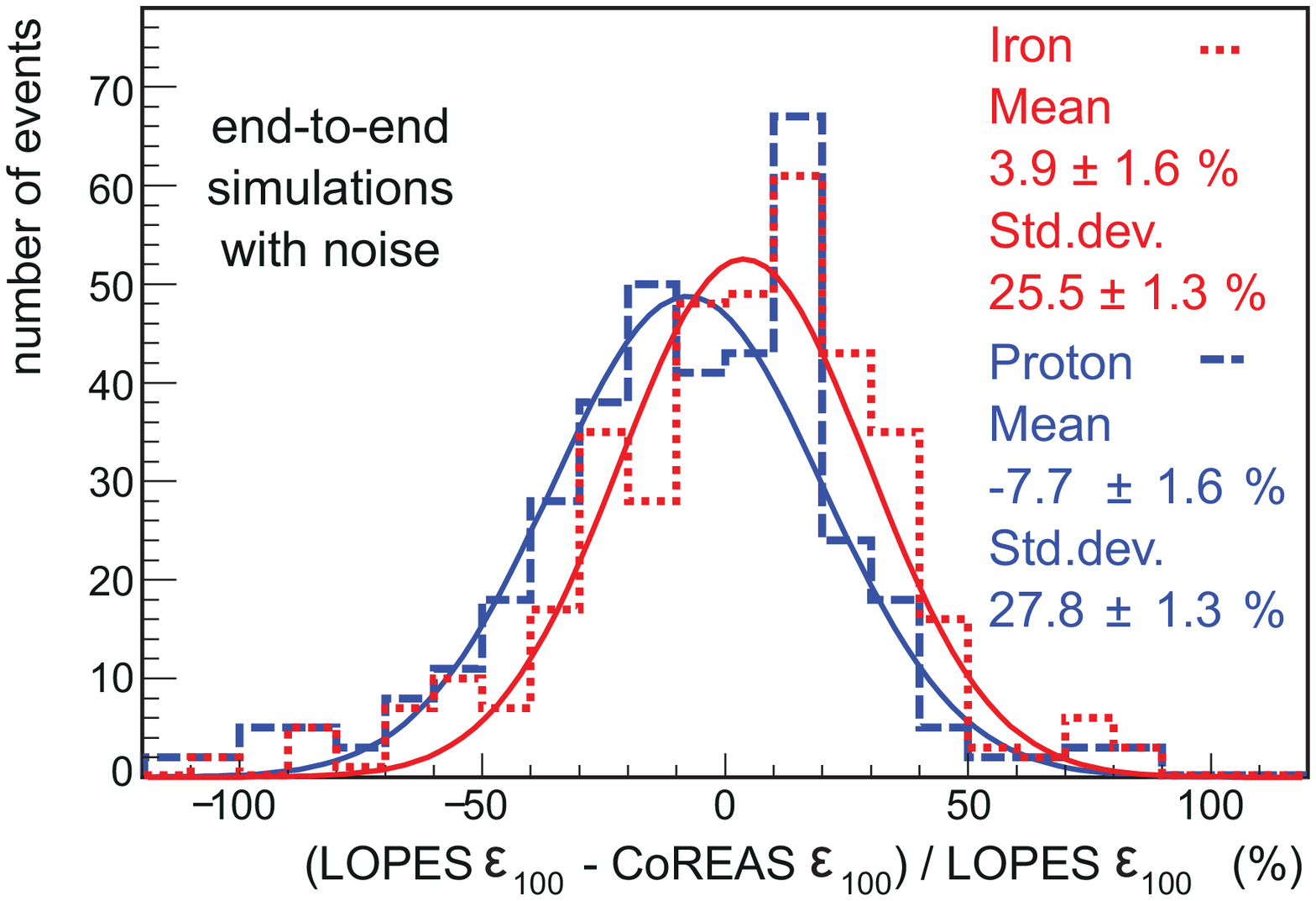}
  \caption{Comparison of LOPES measurements and end-to-end simulations including noise for the cross-correlation amplitude (i.e., the histograms to Fig.~\ref{fig_CCbeamComparison}) and the amplitude $\epsilon_{100}$ at $100\,$m distance from the shower axis as determined by a fit of the lateral distribution (for east-west antennas). 
  The stated values are from a Gaussian fitted to the histograms.
  }
  \label{fig_histAmplitudeComparison}
\end{figure*}

\subsection{Consistency of Measurements and end-to-end Simulations}

Do CoREAS simulations agree with the measurements?
In earlier publications, we had compared the amplitude predicted by CoREAS with the amplitudes measured by LOPES, but were not able to assess the systematic uncertainties due to the simplified treatment of the antenna gain described above.
With the new end-to-end processing of the simulations, we solved this problem and could compare simulations and measurements in an ``apples-to-apples'' way.
The remaining significant uncertainties are the scale uncertainty of the amplitude calibration of $16\,\%$ and the $20\,\%$ uncertainty on the energy reconstructed by KASCADE-Grande that was used as input for the simulations.

We confirmed that CoREAS describes the absolute amplitude well, both for the CC-beam amplitude and for the amplitude $\epsilon_{100}$ at $100\,$m obtained from the LDF fit. 
Determining the offset between simulations and data as a factor $k$ of a line fitted to the event-by-event correlation (Fig.~\ref{fig_CCbeamComparison}) and as the mean of a histogram of the deviations (Fig.~\ref{fig_histAmplitudeComparison}) yielded consistent results.
Also the offsets between simulations and measurements for the CC-beam amplitude and for the amplitude $\epsilon_{100}$ at $100\,$m axis distance are consistent within statistical uncertainties.
The difference of about $11\,\%$ to $12\,\%$ between proton and iron showers can be explained by the different fraction of the energy in the electromagnetic component emitting the radio signal.
In all cases, the mean offset between LOPES and CoREAS is lower than the calibration scale uncertainty.

\begin{figure*}
  \centering
  \includegraphics[width=0.49\linewidth]{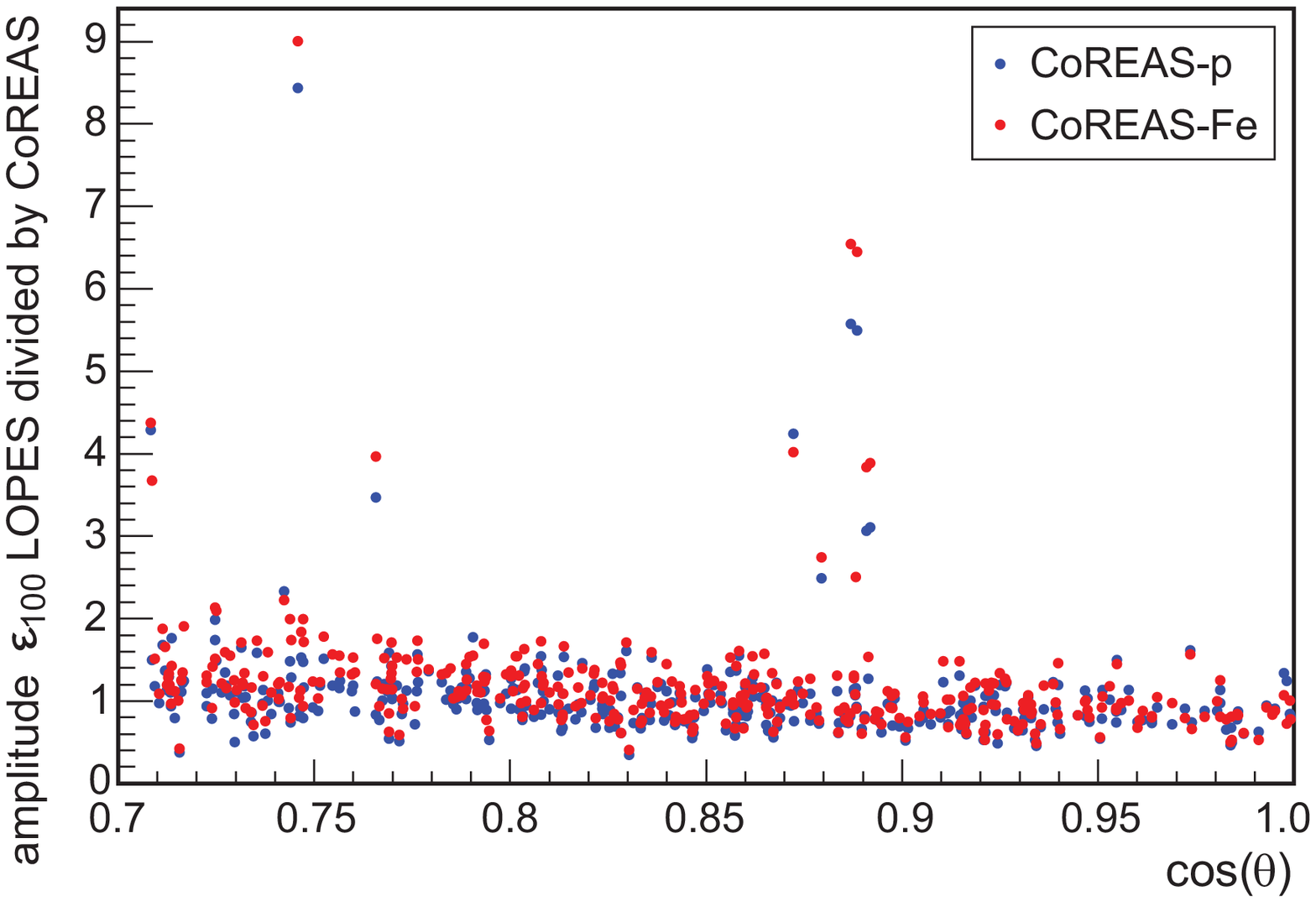}
  \hfill
  \includegraphics[width=0.49\linewidth]{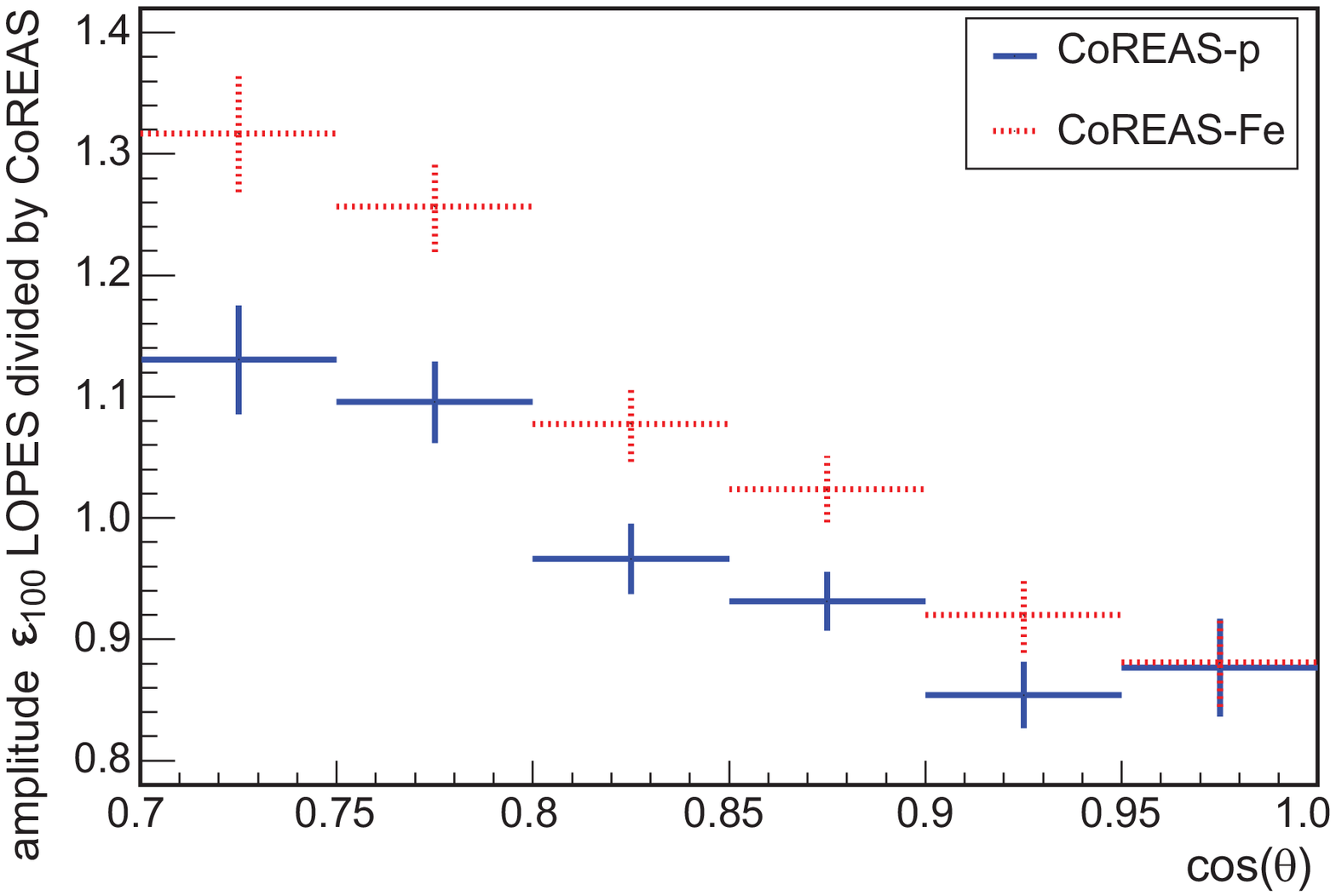}

  \caption{Ratio of the measured amplitude $\epsilon_{100}$ at $100\,$m axis distance and the simulated amplitude of the CoREAS end-to-end simulations including noise for 380 events passing all quality cuts (left). 
  In the profile (right) 11 outliers with ratios larger than 2.5 are excluded (see text for discussion). 
  Deficiencies of the antenna model cannot be excluded and might be the reason for the trend in the ratio versus zenith angle.}
  \label{fig_zenithDependence}
\end{figure*}

Hence, except for a few outliers, the CoREAS simulations are generally consistent with the LOPES measurements. 
These outliers were observed in earlier analyses, too \cite{2013ApelLOPESlateralComparison}. 
For all outliers, the measured amplitude is significantly higher than the simulated one (in the vice-versa situation, an event was likely not detected at all). 
Such upward fluctuation of the measured amplitude could be caused, e.g., by undetected thunderstorms (however, only one of the outliers is an event detected before the thunderstorm monitoring was installed), man-made radio noise overlapping with the air-shower signal, or systematic issues in the instrument response or reconstruction procedure.

A closer investigation of the comparison of the CoREAS simulations and the measurements revealed that there is a trend in the ratio of measured versus simulated amplitudes with zenith angle (Fig.~\ref{fig_zenithDependence}). 
This trend might be due to deficits in the simulated antenna pattern used for the interpretation of the measurements (regular calibration measurements were only done for the zenith, and cannot be repeated for other directions since LOPES is dismantled). 
This unknown reliability of the antenna pattern is a major systematic uncertainty for the interpretation of amplitudes of individual events, but only a smaller uncertainty for average values of the full data set which has its mean at $<\cos \theta> = 0.84$ (with only a smaller difference between the two data sets: $<\cos{\theta}_\mathrm{KASCADE}>=0.86$ and $<\cos{\theta}_\mathrm{Grande}>=0.82$).
The clustering of some of the outliers at a particular zenith angle may be a hint that deficits in the antenna pattern could also be the explanation for some of the outliers.
8 of the 11 outliers in Fig.~\ref{fig_zenithDependence} are also outliers in Fig.~\ref{fig_CCbeamComparison}.
Overall, it seems that there is no unique explanation for all outliers, but a variety of reasons.
In any case, the outliers do not affect the conclusions derived from the average values of the full event statistics presented here.

\begin{figure*}[t]
  \centering
  \includegraphics[width=0.49\linewidth]{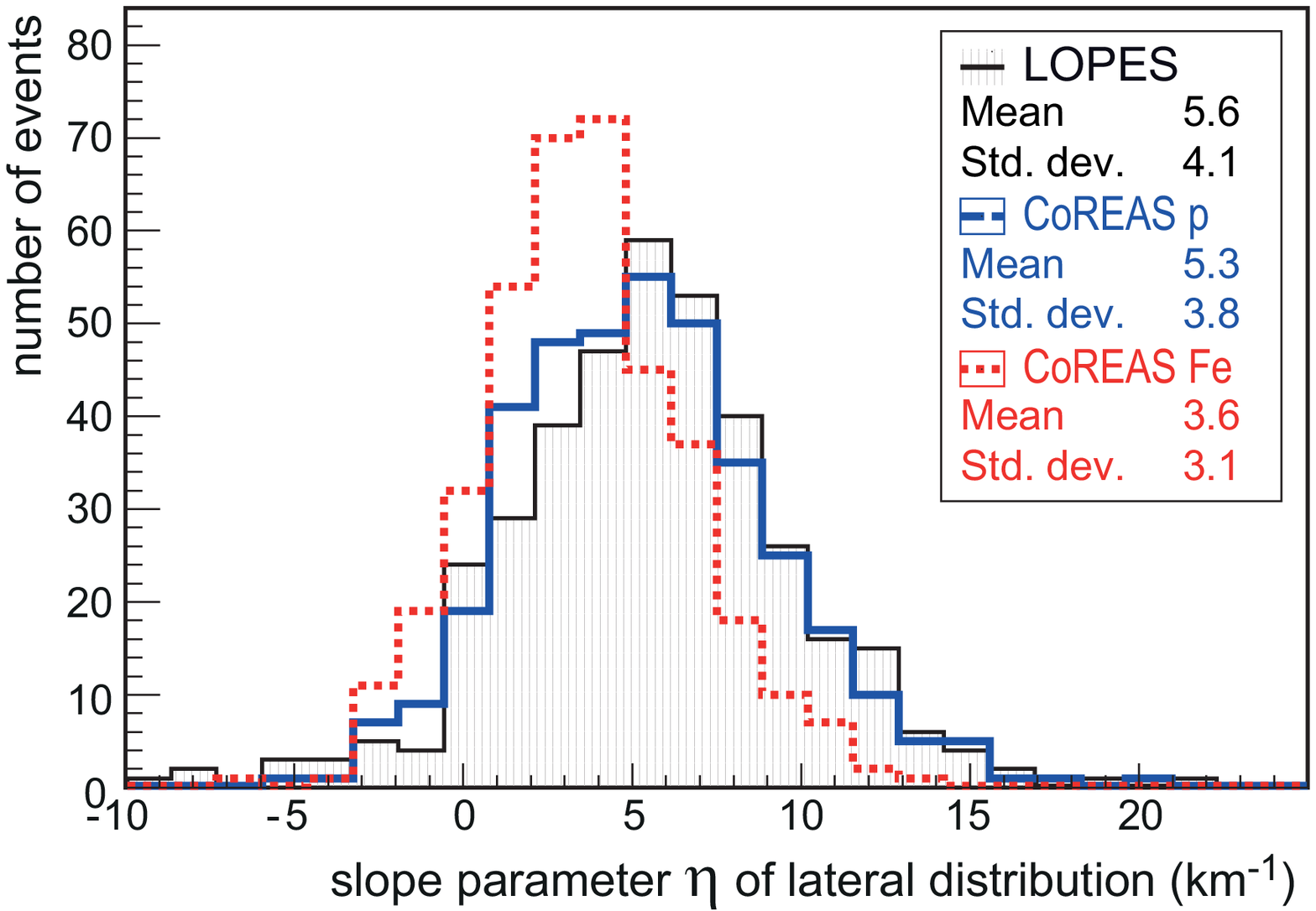}
  \hfill
  \includegraphics[width=0.49\linewidth]{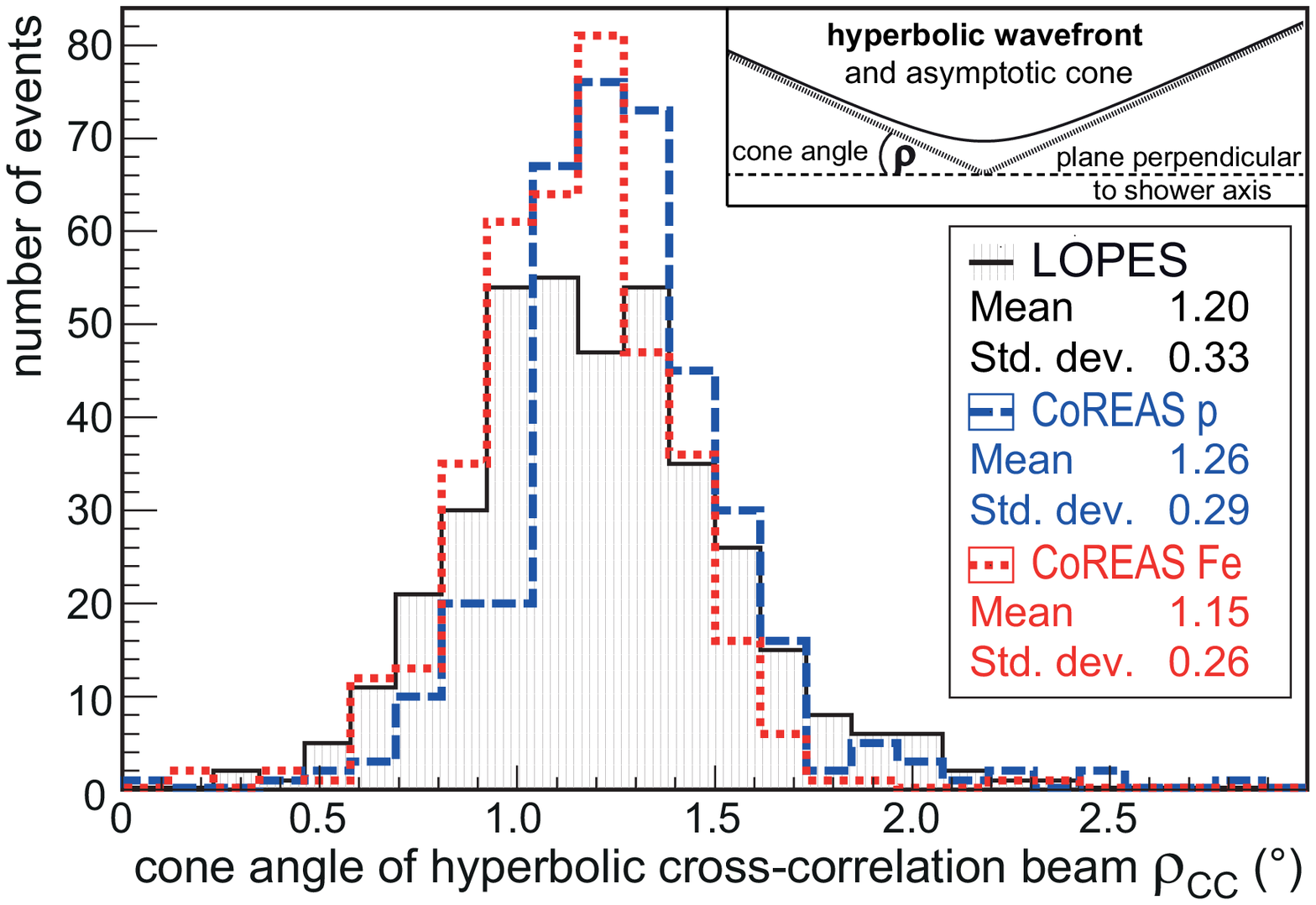}
  \caption{Comparison of LOPES measurements and the end-to-end CoREAS simulations for KASCADE(-Grande) events: 
  the slope of the lateral distribution $\eta$, and the cone angle of the wavefront $\rho_\textbf{CC}$ determined by cross-correlation beamforming.
  }
  \label{fig_RhoAndEtaComparison}
\end{figure*}

In addition to the amplitudes, we also compared higher-level quantities, in particular the slope of the lateral distribution and the shape of the wavefront (Fig.~\ref{fig_RhoAndEtaComparison}). 
The results show that CoREAS simulations are generally compatible with the measurements, though there are some interesting details. 
For both the slope parameter $\eta$ of the LDF and the cone angle $\rho$, i.e, the angle between the shower plane and the asymptotic cone of the hyperbolic wavefront determined by maximizing the CC-beam amplitude, the measured distributions are slightly wider than the simulated ones.
Such a wider distribution is expected if there are additional measurement uncertainties not considered in the end-to-end simulations.
Indeed, there are such uncertainties: the weather-dependent uncertainty on the antenna gain increases the uncertainty on the lateral slope; and occasional glitches and jitters in the time synchronization, though corrected mostly by the beacon, increase the uncertainty on the wavefront reconstruction.
Thus, a slightly wider distribution of the measured parameters than in the end-to-end simulations was expected, and does not mean that CoREAS would not describe the distributions correctly. 
 
Regarding the mean values of the distribution, there is a small inconsistency, which may reflect unknown systematic uncertainties. 
The measured distributions of $\eta$ and $\rho$ are each individually compatible with CoREAS.
However, given that both distributions originate from the same measured events, there is an interesting divergence:
For the slope parameter $\eta$, the measured distribution is close to the proton distribution.
For the cone angle $\rho$, the measured distribution is closer to the iron distribution.
This discrepancy was reported earlier by us \cite{LOPES_ICRC2017}. 
In order to make a conclusion whether there was an unknown systematic effect at LOPES, or whether this discrepancy originates from the simulations, further investigation will be necessary at more accurate interferometric antenna arrays, such as LOFAR \cite{SchellartLOFAR2013} or the SKA \cite{HuegeSKA_ICRC2015}.

\subsection{Relation of the Cross-Correlation Beam and the Lateral Distribution}
The end-to-end simulations finally enabled us to better study the interpretation of the CC-beam amplitude. 
With the first LOPES data we had already shown that the CC-beam amplitude was approximately linearly correlated with the energy of the primary particle after correction for the geomagnetic angle $\alpha$, i.e., the angle between the Earth's magnetic field and the shower axis \cite{FalckeNature2005}.
While the absolute amplitude $\epsilon$ measured in individual antennas is easy to understand, the absolute amplitude of the CC-beam lacked interpretation and it was not clear how the CC-beam amplitude related to the amplitudes in individual antennas.

\begin{figure*}[t]
  \centering
  \includegraphics[width=0.32\linewidth]{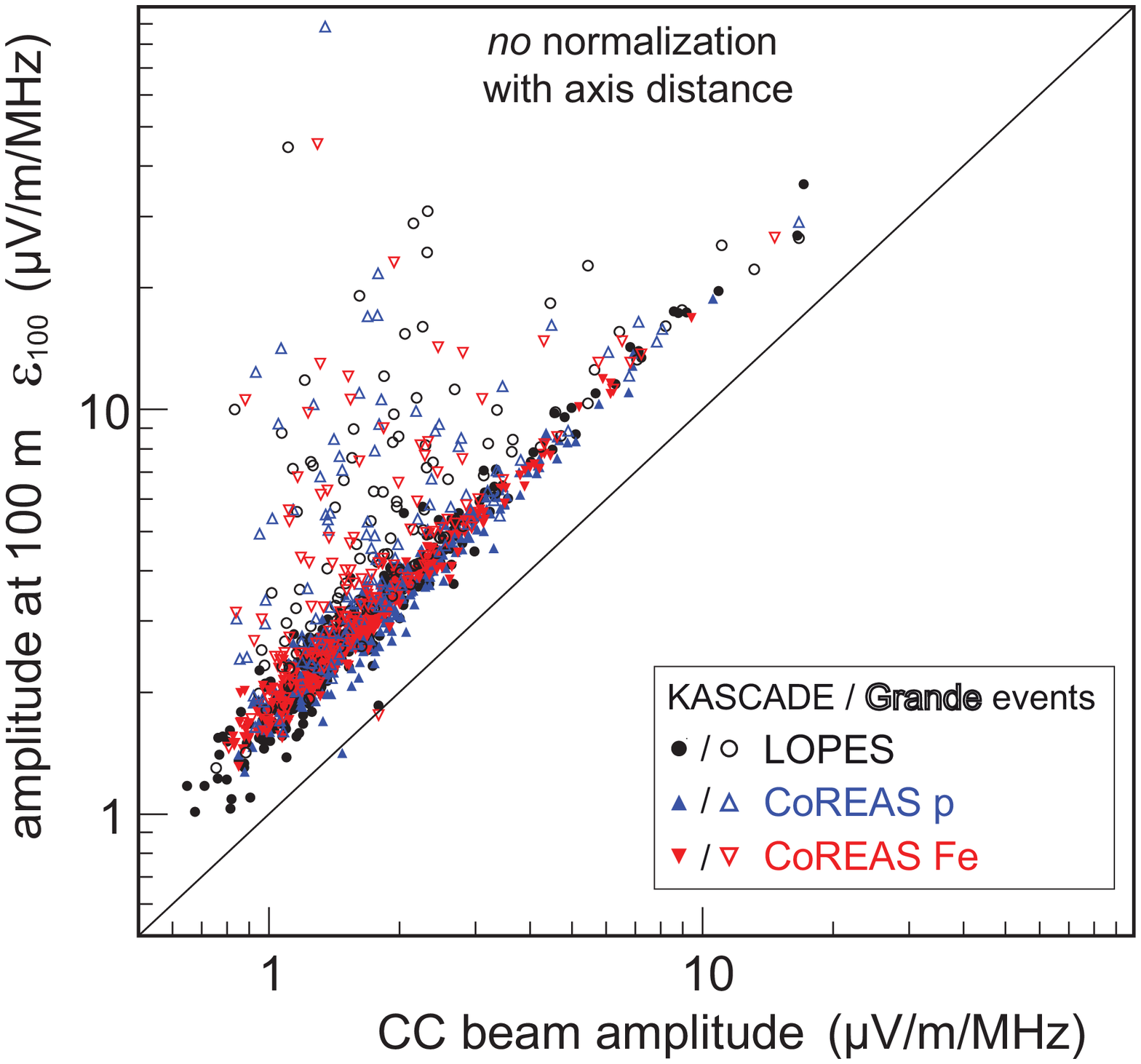}
  \hfill
  \includegraphics[width=0.32\linewidth]{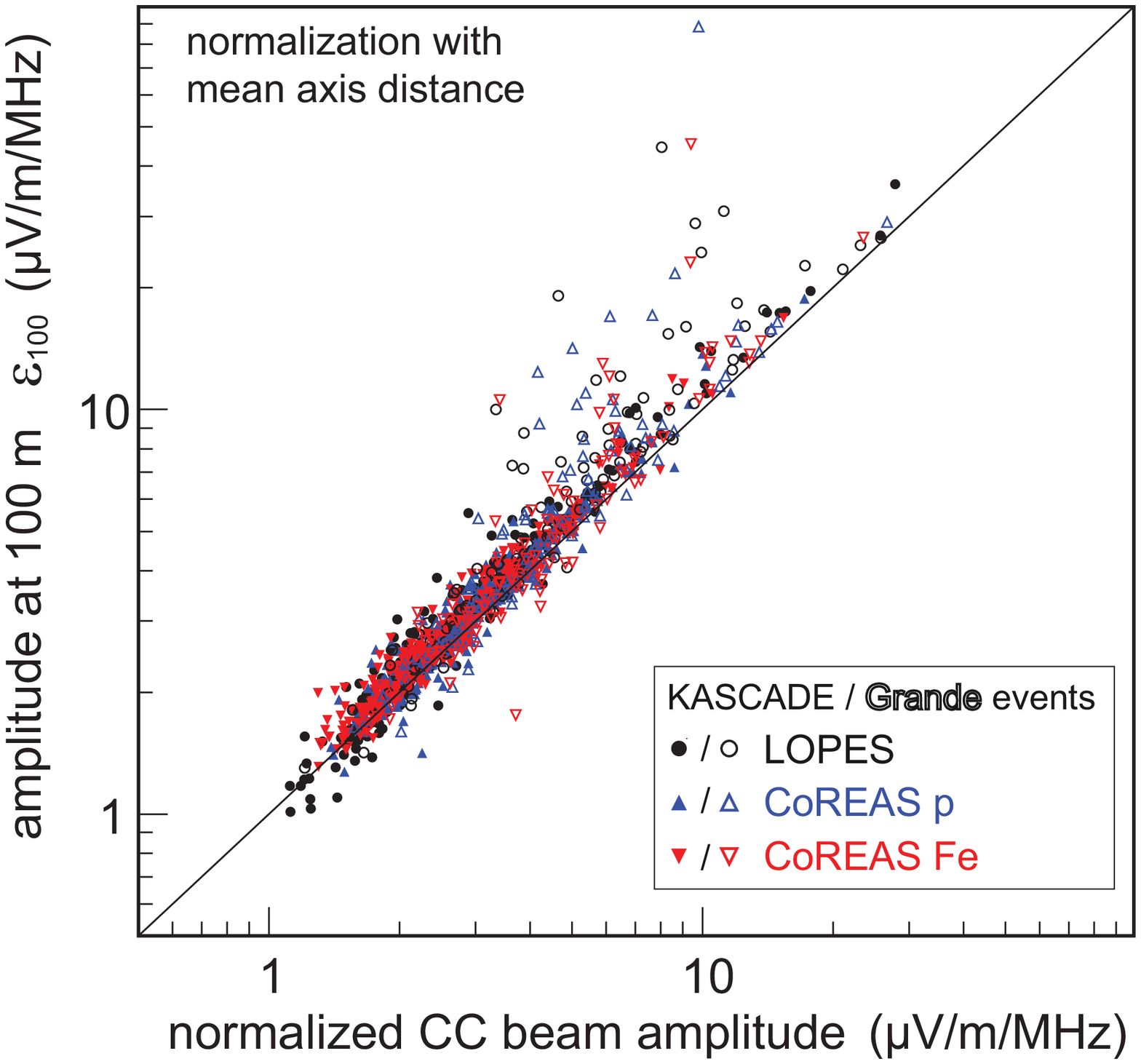}
  \hfill
  \includegraphics[width=0.32\linewidth]{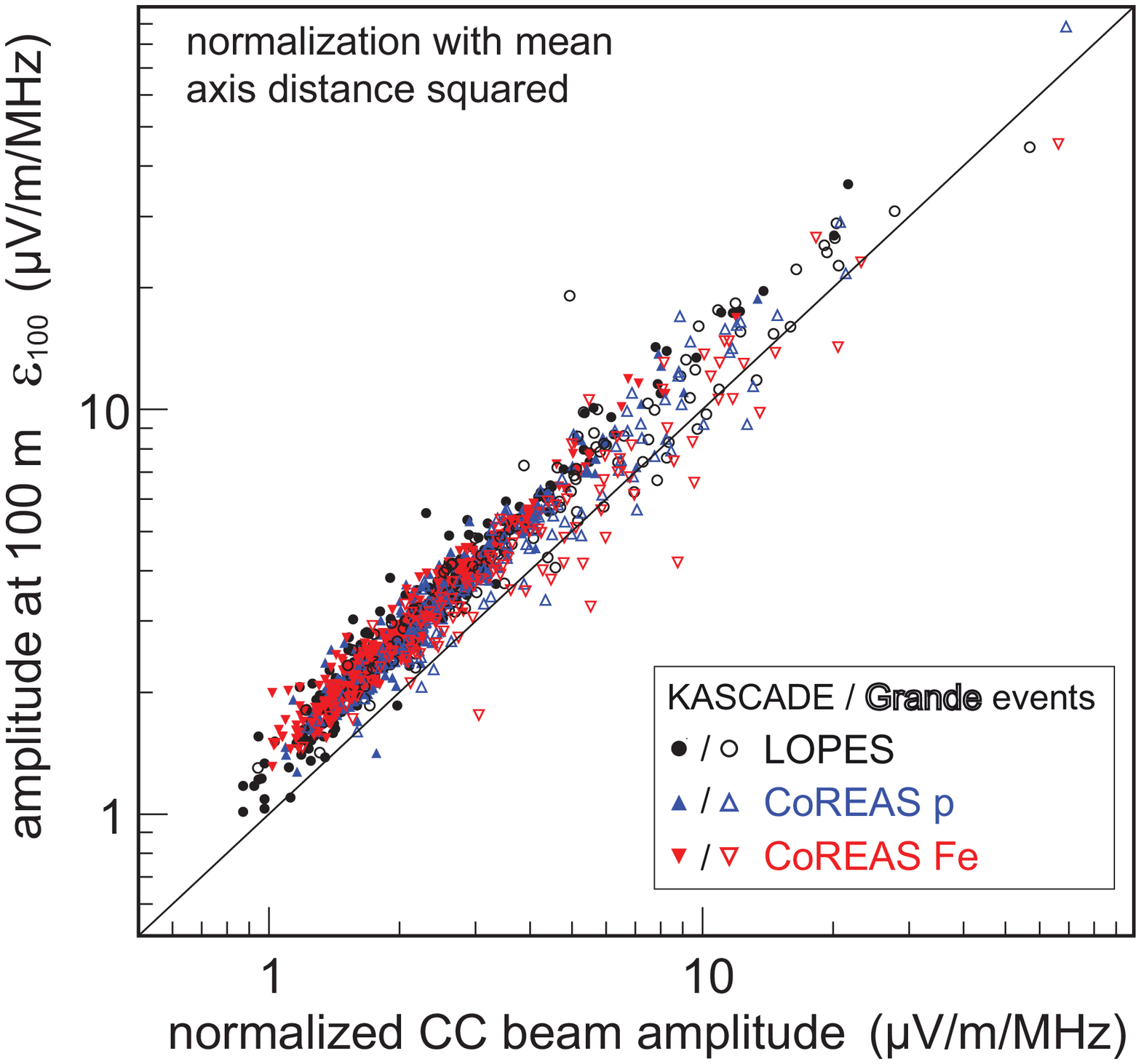}

  \caption{Event-by-event comparison of the amplitude at $100\,$m axis distance and the amplitude of the cross-correlation beam. 
  While for KASCADE events there always is a good correlation, this is not the case for several Grande events (outliers in left and middle plots are Grande events), which have a larger axis distance to the LOPES antennas. 
  Normalizing the CC-beam amplitude by the mean lateral distance improves the correlation significantly.}
  \label{fig_cc_vs_eps100}
\end{figure*}

We now investigated this by comparing the amplitude $\epsilon_{100}$ determined by the LDF with the CC-beam amplitude (Fig.~\ref{fig_cc_vs_eps100}).

We find a linear correlation which has a mean ratio of $<\epsilon_{100}/CC> = 1.83 \pm 0.22$ for the LOPES measurements, $1.79 \pm 0.19$ for the proton simulations, and $1.82 \pm 0.14$ for the iron simulations. 
The stated ratios are the mean and standard deviations of a Gaussian fitted to the ratio of $\epsilon_{100}$ and the (normalized) CC-beam amplitude. 
The outliers are mainly the more distant events, with a core in the Grande array, which show a significantly lower CC-beam amplitude.
Since the CC beam represents a non-trivial way of averaging over the individual antenna signals, this was expected due to the steeply falling lateral distribution of the radio signal.
While the CC beam itself does not contain any correction for the lateral distribution and is expected to depend on the mean distance of the antennas from the shower axis, $\epsilon_{100}$ should be independent of the distance by construction.
The fact that the outliers are consistent in the measurements and end-to-end simulations, indicates that they are not an artifact, but have a physical explanation in the lateral distribution of the radio signal.

One would expect that a normalization of the CC-beam amplitude by the average exponential falloff of the radio amplitude with axis distance would fix the issue. 
Therefore we multiplied all CC-beam amplitudes by $\exp(d_\mathrm{mean}/180\,\mathrm{m})$, where $180\,$m corresponds to the average lateral slope parameter $\eta = 5.6\,$km$^{-1}$ measured with the LDF. 
This normalization improves the correlation and changes the average ratio to $1.09 \pm 0.11$ for the LOPES measurements, $1.06 \pm 0.19$ for the proton simulations, and $1.08 \pm 0.11$ for the iron simulations.
However, for some distant events this correction was insufficient which is visible by the few remaining outliers in the middle plot of Fig.~\ref{fig_cc_vs_eps100}). 
Multiplying the CC-beam amplitude instead by the square of this normalization factor, $\exp(d_\mathrm{mean}/180\,\mathrm{m})^2$, yielded a linear correlation between the CC-beam amplitude and $\epsilon_{100}$ with a ratio of $1.37 \pm 0.16$ for the LOPES measurements, $1.34 \pm 0.15$ for the proton simulations, and $1.36 \pm 0.14$ for the iron simulations.
Note that there is no reason to expect a ratio of $1$; since the amplitude $\epsilon_d$ depends on the reference distance $d$ to the shower axis, choosing a different reference distance will automatically lead to a different value for the ratio. 
Nonetheless, the measurement of this ratio is useful, e.g., for comparing earlier publications using either the CC-beam or the lateral distribution.
Regarding the relative standard deviation, there is no major difference between both normalizations, but there is regarding the outliers which are Grande events at larger distances.
Although the squared normalization is difficult to explain from simple principles, we point out that we observe the improvement regarding the outliers by this normalization consistently in the measurements and simulations. 
This means that albeit empirical, the choice of the normalization is not arbitrary.
It is likely related to the complex interplay of the physics of the radio emission and the detector response, since both are included in the end-to-end simulations.
The consistent effects of these normalizations in the measurements and simulations also show that the CC-beam amplitude can be well studied on an absolute level using the end-to-end simulations.

\begin{figure}
  \centering
  \includegraphics[width=0.99\linewidth]{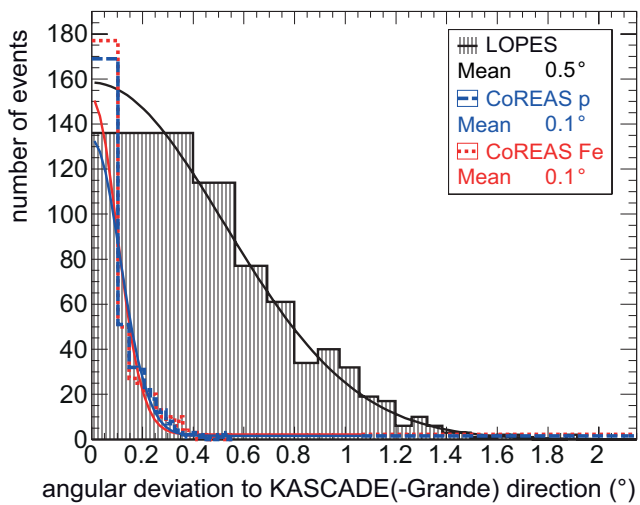}

  \caption{Angular deviation between the arrival direction reconstructed by KASCADE or KASCADE-Grande, respectively, for the LOPES measurements and the CoREAS end-to-end simulations including noise. 
  Simulations without noise (not shown, see Ref.~\cite{LinkPhDThesis2016}) have a mean deviation of $0.05^\circ$. 
  The variable bin size accounts for a constant solid angle covered by each bin.}
  \label{fig_angularDeviation}
\end{figure}

\subsection{Reconstruction of Shower Parameters}

At LOPES we used the cross-correlation beam as well as the pulse time and amplitude in individual antennas to reconstruct the most important shower parameters, which are the arrival direction, the energy, and the depth of the shower maximum ($X_\mathrm{max}$). 
Apart from the improved calibration \cite{2015ApelLOPES_improvedCalibration}, the reconstruction procedure used for the final results presented here had not changed significantly compared to earlier publications. 
However, for the CC beam the reconstruction procedure evolved over time, since we meanwhile used the more accurate hyperbolic wavefront for the beamforming instead of the spherical wavefront used in the early years of LOPES \cite{2014ApelLOPES_wavefront}. 
Moreover, due to the end-to-end simulations, we were able to update the formula for the reconstruction of the primary energy using the CC-beam amplitude. 

The \emph{arrival direction} was reconstructed by maximizing the CC-beam amplitude (cf.~Sec.~\ref{sec_analysisPipeline}). 
With the end-to-end simulations, we checked the intrinsic accuracy by comparing the true and reconstructed directions for each event. 
In simulations without noise, the accuracy is $0.05^\circ$ and in simulations with noise the accuracy is $0.1^\circ$ (Fig.~\ref{fig_angularDeviation}). 
In contrast to particle detectors, Poisson statistics do not limit the accuracy of arrival time measurements of a radio array, which explains why radio detection can easily provide for a high angular resolution.
This high accuracy also shows the importance of an adequate wavefront model, since the deviation between the plane wavefront and the used hyperbolic wavefront is already an order of magnitude larger (the average cone angle measured by LOPES is $1.2^\circ$; see Fig.~\ref{fig_RhoAndEtaComparison}).
Thus, the use of an inadequate wavefront could otherwise dominate the uncertainty of the arrival direction. 

For the LOPES measurements, we can only give an upper limit for the achieved direction accuracy by comparing the directions reconstructed with LOPES and with KASCADE-(Grande). 
Doing so, the average accuracy of LOPES was estimated to be better than $0.5^\circ$.
This value includes the uncertainty of KASCADE-(Grande), which is not known accurately enough to make a reliable estimation of the stand-alone accuracy of LOPES.
It is plausible that the direction accuracy of LOPES was worse than in the simulations because the simulations did not include uncertainties in the relative timing between the antennas and in the core position taken from the KASCADE(-Grande) reconstruction. 
In any case, the resulting upper limit for the accuracy of $0.5^\circ$ is better than needed for most applications in cosmic-ray physics.
This demonstrates that even with a limited size and number of detectors, a radio array can provide for an excellent angular resolution. 

For the reconstruction of the \emph{energy} of the primary particle, we had already shown that in addition to the amplitude of the lateral distribution at a specific distance \cite{2014ApelLOPES_MassComposition} also the CC-beam amplitude provides an energy estimator \cite{FalckeNature2005}. 
Our new results confirm that the CC-beam method features a measurement accuracy approximately equal to the lateral-distribution method. 
With the end-to-end simulations, we determined an independent absolute calibration for the energy reconstructed using the CC beam. 
Furthermore, we updated the way how we normalize to the angle between the geomagnetic field and the shower axis and to the mean axis distance of the antennas contributing to the CC-beam - resulting in the following formula for the reconstruction of the energy $E$ of the primary particle:  




\begin{figure*}[t]
  \centering
  \includegraphics[width=0.4\linewidth]{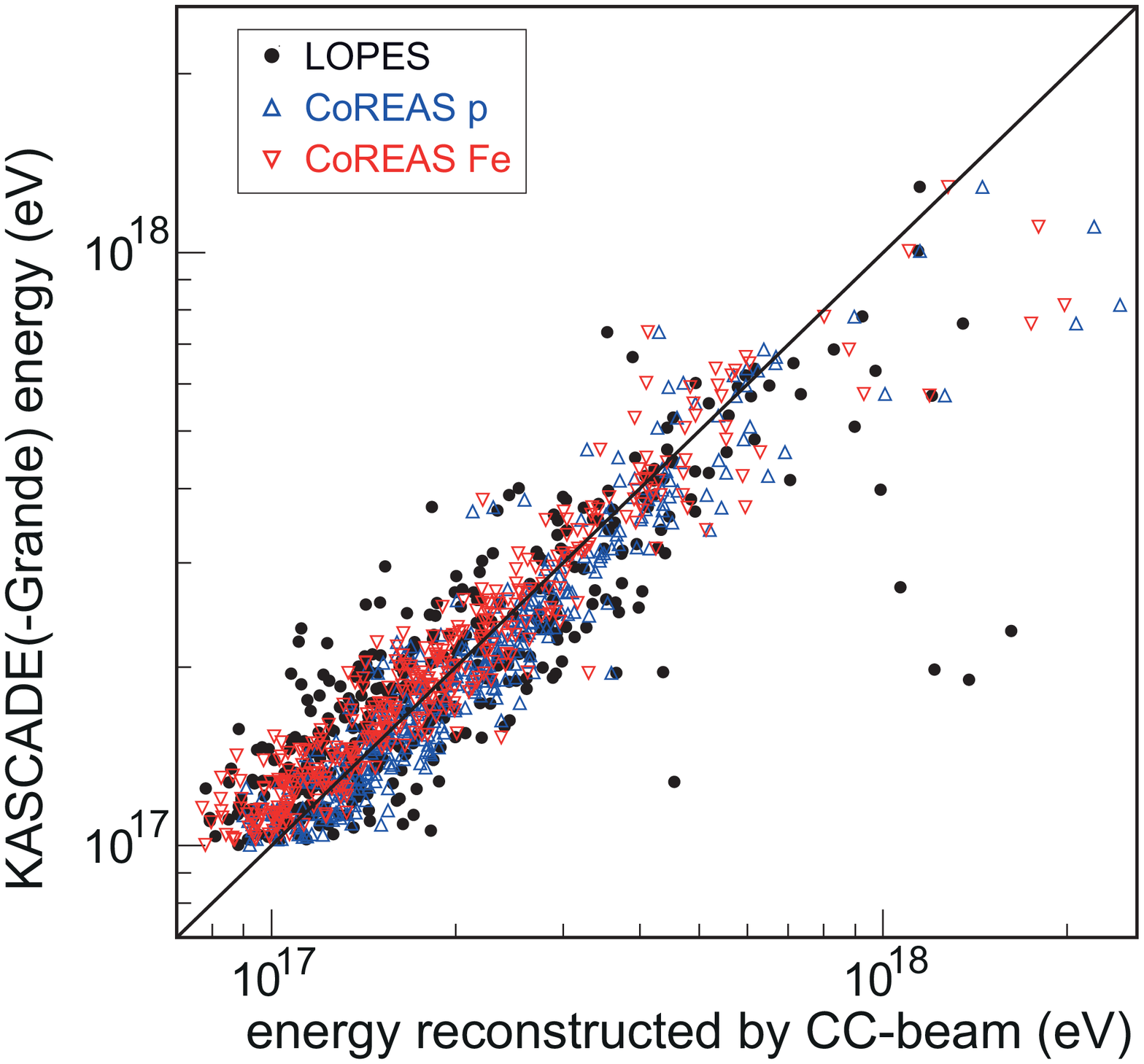}
  \hfill
  \includegraphics[width=0.49\linewidth]{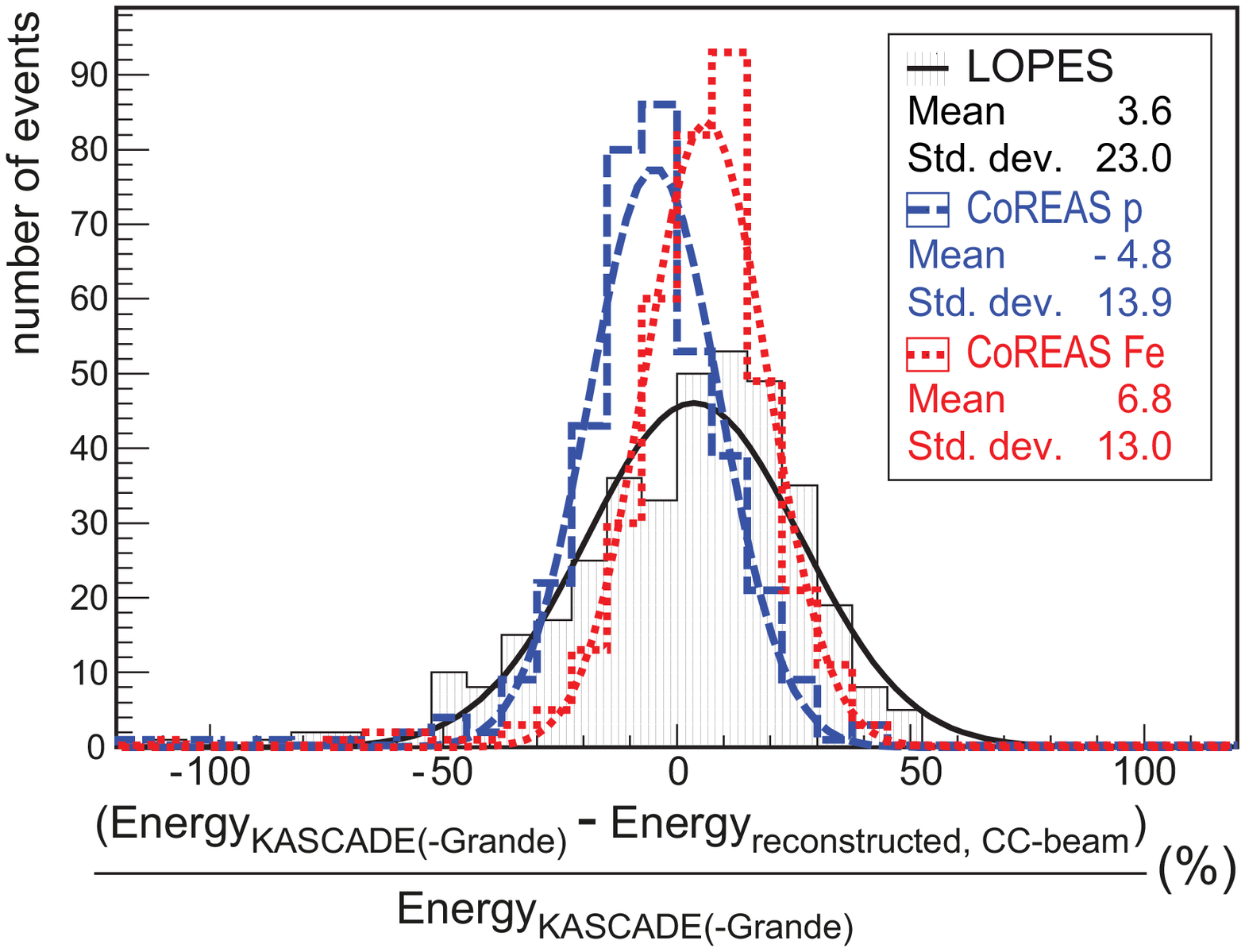}
  \caption{Comparison of the energy reconstructed by the amplitude of the cross-correlation beam to the KASCADE(-Grande) energy for LOPES measurements and CoREAS simulations including noise.
  Left: Per-event comparison. The outliers in the upper-right corner visible also in the simulations have a small east-west component of the geomagnetic Lorentz vector of $p_\mathrm{EW} \le 0.1$; for two of them the corresponding LOPES points are at energies about three times higher than the plotted range; for the other outliers see the discussion related to Figs.~\ref{fig_CCbeamComparison} and \ref{fig_zenithDependence}.
  Right: Histogram of the relative deviation, where the standard deviation provides a measure of the energy precision.
  }
  \label{fig_CCbeamEnergy}
\end{figure*}

\begin{equation}
 E = \kappa \cdot \frac{CC \cdot\exp{(d_\mathrm{mean}/180\,\mathrm{m}) }}{|\vec{v} \times \vec{B}|_{EW}}
 \label{eq_energy}
\end{equation}

with $CC$ the amplitude of the cross-correlation beam, $d_\mathrm{mean}$ the mean axis distance of the antennas contributing to the measurement of the event, and $\vec{v} \times \vec{B}$ the unit vector of the geomagnetic Lorentz force, i.e., we normalized the CC-beam amplitude by the size of its east-west component, and $\kappa$ a proportionality factor determined by the average of the proton and iron simulations \cite{LinkPhDThesis2016}: for the 258 KASCADE events the used proportionality factor is $\kappa = 41.9\,$PeV/(\textmu V/m/MHz) and for the 122 Grande events $\kappa = 36.7\,$PeV/(\textmu V/m/MHz).
In both cases the values were determined as mean of the proton and iron simulations because showers initiated by heavy primaries have a slightly lower radio amplitude on ground than those initiated by light particles (cf.~Fig.~\ref{fig_histAmplitudeComparison}). 
By this choice of the mean value, the maximum energy bias for an individual event is $\pm 6\,\%$ which is small compared to the total energy uncertainty.
In contrast to the observation in Fig.~\ref{fig_cc_vs_eps100}, the normalization by $\exp(d_\mathrm{mean}/180\,\mathrm{m})^2$ instead of $\exp(d_\mathrm{mean}/180\,\mathrm{m})$ deteriorated the correlation of the CC-beam amplitude with the energy instead of improving it (this again is an empirical observation which is consistent in the LOPES measurements and the CoREAS end-to-end simulations).

We have investigated possible reasons for the need of two different proportionality factors $\kappa$ for the KASCADE and Grande events:
A possible explanation is that KASCADE events are mostly contained in the array, while the core of Grande events generally is outside the LOPES array. 
Moreover, there might be a selection bias, since only a small fraction of the triggered Grande events is detected by LOPES.
However, neither simple upward fluctuations nor systematic measurement uncertainties are possible causes of the effect because the factors $\kappa$ have been derived from the end-to-end simulations. 
Nonetheless, also the LOPES measurement of the energy is systematically higher for Grande events than for KASCADE events when using an equal proportionality factor as a cross-check. 
The effect seems to be even stronger in the measurements, probably due to systematic uncertainties or selection biases not present in the simulations, such as upward fluctuations of the CC beam increasing the detection probability. 
However, a solid quantitative investigation is not possible because most LOPES events do not fulfill the minimum quality criteria for the KASCADE and Grande energy measurements simultaneously.
Consequently, while a firm conclusion is difficult due to the limitations of the experimental setup, it appears that simulations and measurements are at least qualitatively consistent regarding the need of different proportionality factors.

We checked the accuracy of the energy reconstruction by Eq.~\ref{eq_energy} by comparing the true and reconstructed values for the end-to-end simulations with noise, and by comparing the LOPES and KASCADE-Grande reconstruction for the measured events (Fig.~\ref{fig_CCbeamEnergy}). 
According to the simulations, the energy precision could be as good as $13\,\%$ to $14\,\%$, with an additional bias of the order of $\pm 6\,\%$ if the mass of the primary particle is unknown. 
This would result in a total average accuracy of about $15\,\%$ for the case of a mixed, unknown mass composition. 

For the LOPES measurements, the mean deviation to the KASCADE(-Grande) reconstruction is $23\,\%$. 
This includes the KASCADE(-Grande) uncertainty of the energy, which can explain most of the difference to the accuracy expected from the end-to-end simulations.
Additionally, the larger mean deviation indicates that the measurements include additional uncertainties not considered in the end-to-end simulations, such as the deficiencies in the antenna model. 
On the one hand, these deficiencies contribute significantly to the systematic energy uncertainty of individual events, and will increase the spread of the distribution. On the other hand, the effect of the antenna model on average values over a larger data set is expected to be small compared to the $16\,\%$ scale uncertainty of the amplitude calibration, which is the largest systematic uncertainty.

We emphasize, that in contrast to earlier analyses, we did not tune the energy formula to the KASCADE(-Grande) measurements, but instead to the CoREAS simulations. 
Therefore, the fact that the mean offset between the LOPES and KASCADE(-Grande) energy measurements is smaller than the $16\,\%$ scale uncertainty of the absolute amplitude calibration confirms again that the CoREAS simulations are compatible with the LOPES measurements. 
In addition, all simulations depend on the hadronic interaction models in use.
This might affect the reconstruction of KASCADE-Grande measurements more than the reconstruction of LOPES measurements since the radio emission is sensitive to the well understood electromagnetic shower component, only.  
For both, the simulations and measurements, the accuracies achieved by the CC beam for the energy are similar to that achieved by using the amplitude at a specific lateral distance \cite{2014ApelLOPES_MassComposition}.
Consequently, for other experiments it might be worthwhile to investigate both methods.

\begin{figure*}[t]
  \centering
  \includegraphics[width=0.49\linewidth]{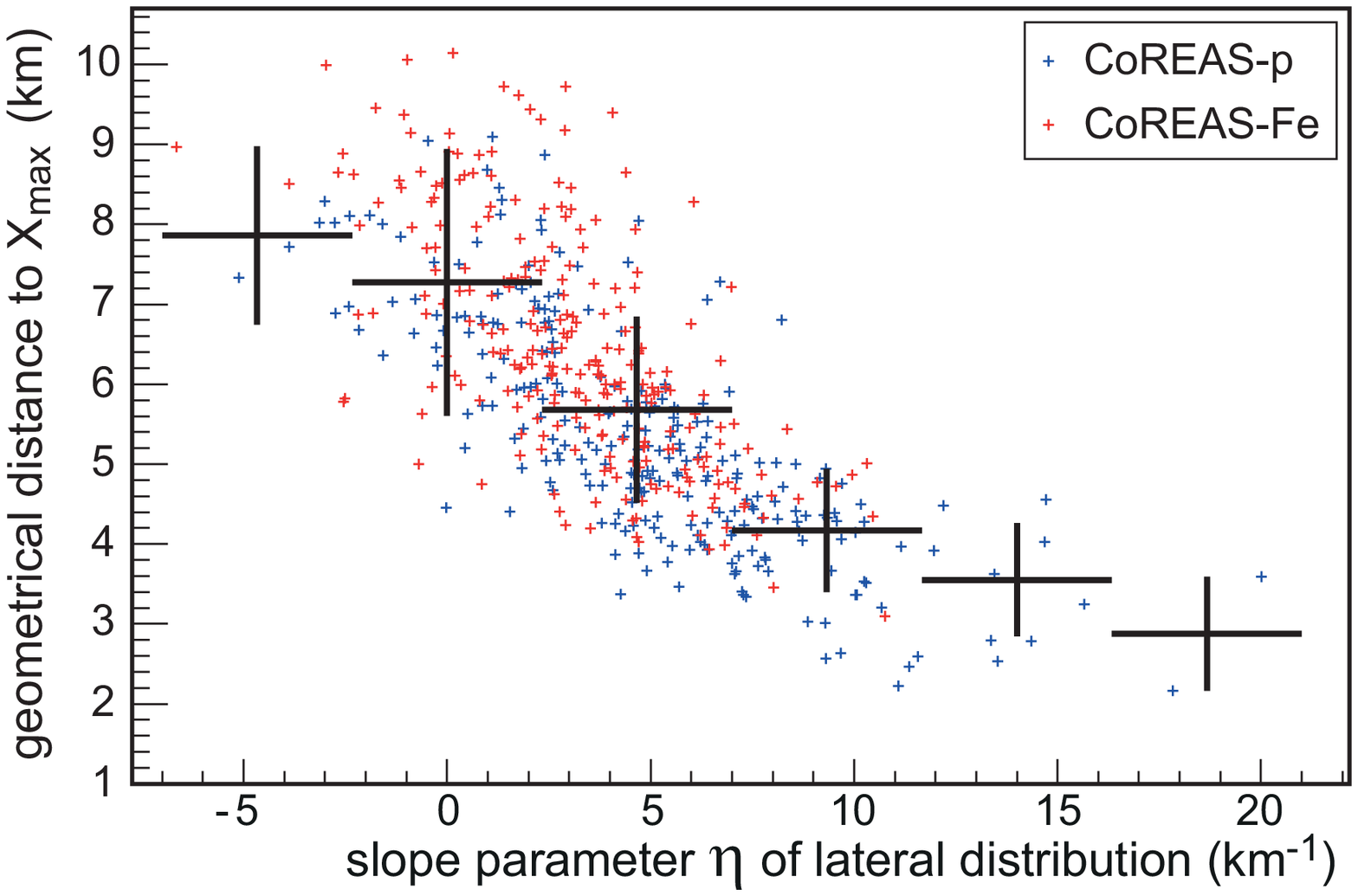}
  \hfill
  \includegraphics[width=0.49\linewidth]{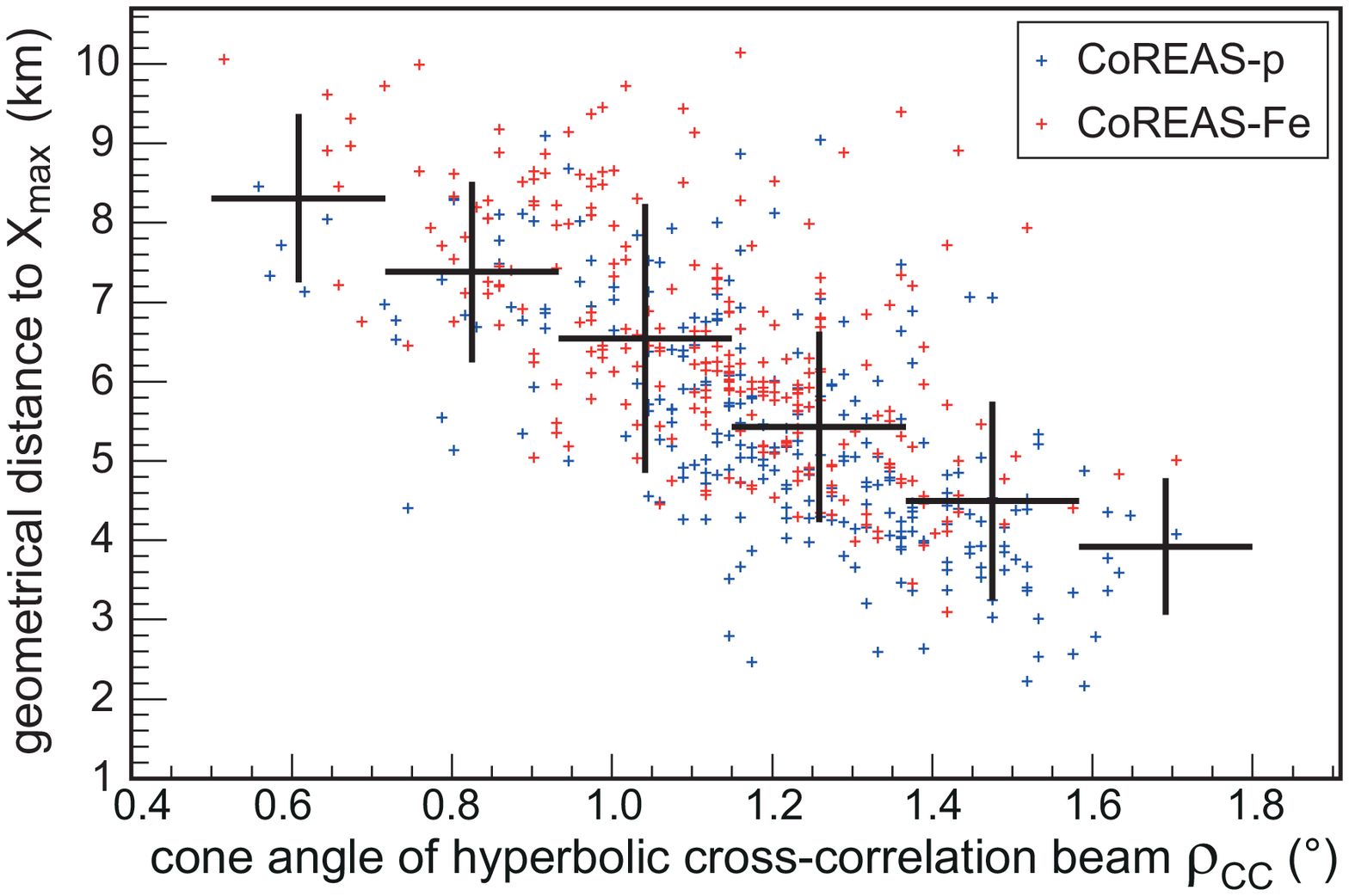}
  \caption{Correlation of the lateral slope parameter $\eta$ (left) and of the wavefront cone angle $\rho$ (right) of the CoREAS end-to-end simulations (with noise) with the geometrical distance to the shower maximum for the contained KASCADE events. The profile in black denotes the mean and standard deviation in each bin.
  }
  \label{fig_Xmax}
\end{figure*}

As experimentally demonstrated by LOPES, radio measurements on ground are sensitive to the longitudinal shower development \cite{2012ApelLOPES_MTD}, and therefore can be used for the reconstruction of the position of the \emph{shower maximum}. 
The two methods used by LOPES are based on the slope of the lateral distribution \cite{2014ApelLOPES_MassComposition} and on the opening angle of the asymptotic cone of the hyperbolic wavefront \cite{2014ApelLOPES_wavefront}. 
The more distant the shower maximum from the array, the flatter is the lateral distribution and the flatter is the wavefront. 
With the end-to-end simulations we confirm our earlier results that both parameters are statistically correlated with the distance to the shower maximum (Fig.~\ref{fig_Xmax}).
In earlier publications \cite{LOPES_ICRC2017}, we had used these dependencies to determine the mean atmospheric depth of the shower maximum, $X_\mathrm{max}$, though with limited precision.
As at the time of the original publications, we are unable to quantify the additional systematic uncertainties, e.g., the wavefront reconstruction seems to be sensitive to the uncertainty of the core position.
The wavefront method might therefore be useful primarily for dense arrays with a good resolution of the shower core, such as LOFAR or the SKA.

As noted earlier \cite{2014ApelLOPES_wavefront,LOPES_ICRC2017}, there is a difference in the absolute scale of $X_\mathrm{max}$ resulting from the wavefront and the lateral slope methods. 
While both methods show a sensitivity to $X_\mathrm{max}$, i.e,. are suitable to measure $X_\mathrm{max}$ of an individual event with a certain precision, there seems to be an additional uncertainty on the absolute scale depending on the method. 
Understanding and quantifying these scale uncertainties will be crucial for an accurate interpretation of $X_\mathrm{max}$ in terms of mass composition. 
For this purpose, we suggest that also other experiments could compare their $X_\mathrm{max}$ reconstructions of an amplitude-based (e.g., lateral slope) and a timing-based (e.g., wavefront) method.
This will be most important for those experiments using radio as only technique for $X_\mathrm{max}$, but will only be a minor issue for hybrid arrays also featuring optical detectors available for cross-calibration of the scale \cite{TunkaRex_Xmax2016,AERA_XmaxMethods_ARENA2016}.
Moreover, better methods for the estimation of the efficiency and aperture are needed to avoid a bias caused by the partial efficiency of a radio array close to its threshold \cite{LenokICRC2019}.

\section{Conclusion}
LOPES was a successful pathfinder for the radio detection of air showers in the digital era. 
It was a driver and initiator of the meanwhile matured detection technique for ultra-high-energy cosmic rays, and provided essential contributions to the understanding of the radio emission and the potential of the detection technique. 
While today there is a general agreement in the community on the principles of the radio emission, this was not the case when LOPES started, and its measurements helped to understand deficits in earlier models and simulation codes.
Now CoREAS simulations and LOPES measurements are compatible within uncertainties for all tested parameters, which shows that the physical processes of the radio emission are well understood. 
Even so, the question remains why the wavefront and lateral slope methods result in a slightly different $X_\mathrm{max}$ scale although both methods were calibrated using the same CoREAS simulations. 
Further investigations at other experiments will help to investigate whether this is due limitations of our understanding of the LOPES detector or due to the simulations.

Independent of the simulations, the comparison of the LOPES and KASCADE-(Grande) measurements provided the proof-of-principle that the radio technique can be used for accurate measurements of the arrival direction and energy of cosmic rays -- even in a noisy environment. 
Using the full potential of the radio technique, including an accurate reconstruction of $X_\mathrm{max}$, however, requires a location more radio-quiet than the LOPES site. 
This has meanwhile been demonstrated by several second-generation experiments following LOPES \cite{LOFARNature2016,TunkaRexXmaxPRD2018,AERA_XmaxMethods_ARENA2016}.

LOPES also pioneered many methods regarding the radio technique which have been used by other experiments, such as LOFAR, AERA, and Tunka-Rex.
Examples are the continuous monitoring of the atmospheric electric field and the calibration methods used by LOPES, in particular the monitoring of the relative timing with a beacon \cite{SchroederTunkaRex_PISA2015,AERAairplanePaper2015} and the end-to-end amplitude calibration using an absolutely calibrated external reference source \cite{TunkaRex_NIM_2015,AERAantennaCalibration2017,NellesLOFAR_calibration2015}.
The latter also enabled the comparison of the absolute energy scales of different air-shower arrays using radio measurements \cite{TunkaRexScale2016}. 
Finally, the public availability of the data on the KCDC platform sustains the LOPES measurements for future analyses.  

So what questions remain? 
The most important one is about the relevance of the technique for the progress of the general field. 
Can the radio technique alone or in combination with particle-detector arrays provide new knowledge on cosmic-ray physics or the particle physics of air-showers?

As stand-alone technique, radio antennas can be used to instrument huge areas for a reasonable price, as planned for the GRAND array aiming at ultra-high-energy cosmic rays and neutrinos \cite{GRAND_WhitePaper2020}. 
For the solution of many of the open questions in high-energy astroparticle physics, a major increase in measurement accuracy for the properties of the primary particle is required, in particular for its mass \cite{SchroderAstro2020,SarazinAstro2020}.
The radio technique now seems to deliver a measurement accuracy mostly equal to the established techniques, but can it also enhance the total accuracy over the state-of-the-art?

The results of LOPES and other experiments already provided hints that antenna arrays can improve the energy and direction accuracy, and possibly also the accuracy of $X_\mathrm{max}$ providing access to the elemental composition of cosmic rays. 
These gradual improvements of the state-of-the-art are important. 
Yet, there are two ideas under investigation that go beyond and may provide for major breakthroughs with the next generation of antenna arrays:
\begin{itemize}
\item Can a significant increase in antenna density increase the overall accuracy for air showers and possibly reveal substructures in the shower development? 
This remains to be shown by LOFAR or the SKA \cite{HuegeSKA_ICRC2015}.
\item An alternative idea is to combine radio and muon detectors in hybrid arrays because the radio and muon signals contain complementary information on the shower development \cite{FalckeNature2005}.
As supported by a recent simulation study \cite{HoltEPJ_C}, the combination of radio and muon detectors can indeed provide for an increase of the accuracy for the mass composition of cosmic rays.
This idea still needs deeper investigation and experimental demonstration at appropriate air-showers arrays, such as the planned enhancement of IceTop \cite{HaungsIceCubeSurfaceUHECR2018,SchroederICRC2019_SouthPoleRadio} and the AugerPrime upgrade of the Pierre Auger Observatory \cite{HorandelAugerRadioUpgradeUHECR2018,PontICRC2019_AugerRadioUpgrade}.
\end{itemize}

In summary, the LOPES results, experiences, and lessons learned remain a valuable resource for the current and next generation of digital antenna arrays for air-shower detection.
These are now dedicated to all kinds of cosmic messengers -- high-energy cosmic rays, photons, neutrinos -- and it becomes likely that radio arrays will play a major role in high-energy astroparticle physics during the coming decades.

\begin{acknowledgements}
We thank the reviewers for carefully checking the manuscript. Their suggestions helped to improve the paper.
LOPES and KASCADE-Grande were supported by the German Federal Ministry of Education and Research. KASCADE-Grande was partly supported by the MIUR and INAF of Italy, the Polish Ministry of Science and Higher Education and by the Romanian Authority for Scientific Research UEFISCDI (PN19060102 and PN19150201/16N/2019 grants). The present study was also supported by grant VH-NG-413 of the Helmholtz association and by the ’Helmholtz Alliance for Astroparticle Physics - HAP’ funded by the Initiative and Networking Fund of the Helmholtz Association, Germany. 
\end{acknowledgements}


\appendix
\section*{Appendix - Lessons Learned}
LOPES being among the first digital radio arrays for air showers, we also used it as a test facility for various new technologies and ideas related to radio detection of cosmic rays. 
Not all of them were published in refereed journals and not all of them were successful. 
Nonetheless, there are lessons learned form these approaches and the design and data analysis of future experiments may profit from the experience made.
For these reasons, we give a summary of these approaches and their results and other lessons learned.

\section{Self-trigger}
LOPES was successful due to its external trigger by KASCADE(-Grande). 
While a self-trigger might be beneficial for some science cases, it is not critical when radio is added as additional technique to a hybrid array featuring particle detectors for the purpose of increasing the overall measurement accuracy for cosmic rays.
However, for very inclined showers in addition to the radio signal only muons arrive at ground, and the low particle density provides a challenge for particle detectors. 
Thus, there are other experimental approaches such as ANITA \cite{ANITA_CR_PRL_2010} and GRAND \cite{GRAND_WhitePaper2020} which require stand-alone radio detection with a self-trigger. 

At LOPES we developed hardware and algorithms for self-triggering at the LOPES\textsuperscript{STAR} setup \cite{SchmidtIcrc2009}, which consisted of $10$ additional logarithmic periodic dipole antennas in the KASCADE-Grande array. 
These efforts led to effective techniques for filtering RFI and rejecting background pulses. 
Although the algorithms were successful to identify air-shower pulses in previously recorded test samples, we were not successful in self-triggering events with the deployed hardware. 
One of the problems was that also anthropogenic background pulses often led to coincident signals in several antennas. 
Nonetheless, several experiments have meanwhile successfully demonstrated self-triggering at more radio quiet sites \cite{ANITA_CR_PRL_2010,RAugerSelfTrigger2012,Kelley:2013rma,Revenu:2019imx,ARIANNA_2017,TREND2019}.

\section{Measurements at $50-500\,$kHz}
Motivated by historic measurements, we also tried to detect air showers at much lower frequencies in the band of $50-500\,$kHz. 
Using the standard KASCADE trigger provided for LOPES, we did not detect any air shower in this band in $13$ days of measurements although more than $70$ of the triggering showers had energies above $10^{17}\,$eV.
The sensitivity was likely limited by the radio-loud environment of LOPES.
Hence, we derived an upper limit of $136^{+48}_{-47}\,$mV\,m$^{-1}$\,MHz$^{-1}$ for the field strength of the vertical component of the radio signal emitted by the air showers in this frequency band \cite{LinkDiplomaThesis2006}.

\section{Tripole Antennas}
In the last stage of LOPES, called LOPES-3D, ten 'tripole' antennas were deployed, each consisting of three orthogonally aligned dipoles in east-west, north-south, and vertical direction \cite{2012ApelLOPES3D}. 
Due to the very high background level for the vertical polarization, only few events with signal in all three polarization channels were detected \cite{HuberPhDThesis2014}. 
The measured polarization of these events confirmed the prediction for geomagnetic emission. 

One of the motivations of deploying a third polarization channel was that the signal would be over-determined if the arrival direction was known. 
This means that in theory the arrival direction of the radio signal could be reconstructed from the measurement of a single tripole antenna. 
Due to the specific polarization pattern of the radio emission on ground caused by the interplay of the Askaryan and geomagnetic emission mechanisms, a sufficiently accurate measurement of the electric field vector at each antenna position can also be utilized for many other purposes, such as refining the position of the shower core or rejecting RFI \cite{Huege:2019ufo}.
Another motivation was to test whether the threshold for very inclined events could be lowered by the vertical channel, since by simple geometry considerations the radio emission of inclined air showers features a significant vertical polarization component. 
At LOPES-3D we were not able to demonstrate either of this, and it is not clear whether the level of vertically polarized background would be low enough at other locations to successfully apply one or both ideas. 
If measurements with three polarization were repeated elsewhere, we suggest considering to rotate the whole tripole setup by $45^\circ$ towards the horizontal plane as already done by others \cite{2005AIPC..745..770G,Revenu:2017rui}.
This would avoid one explicit vertical polarization channel, and make the antenna setup easier to understand due to the rotational symmetry around the pole where the antenna is mounted.

\section{Interferometry of Air Showers}
At LOPES we successfully applied cross-correlation beamforming as an interferometric method to lower the detection threshold in a radio-loud environment.
The reasoning behind CC beamforming is that the signal has the same time structure in all antennas, while any background or noise have not. 
Hence, CC beamforming enhances the signal-to-noise ratio with increasing number of antennas. 
This technique is widely used in radio astronomy: a signal arriving as plane wave from distant sources can be assumed to be the same in all antennas.
 
For air showers the radio signal changes significantly from antenna to antenna, and it is a priori not clear that cross-correlation beamforming is useful. 
Nevertheless, cross-correlation beamforming turned out to be a key asset of LOPES for identifying air-shower pulses against the background. 
Probably because the measured pulse shape was primarily determined by the filter response and not by the original pulse shape emitted by the air showers, the signal structure was similar enough in the individual antennas, and the signal-to-noise ratio was enhanced even though the signal was not equal in all antennas. 

Nonetheless, these complications make the absolute value of the CC amplitude difficult to interpret and to compare between experiments. 
For the same radio footprint at ground, a different antenna array might measure different CC-beam amplitudes.
Consequently, the application of CC beamforming or other interferometric methods needs to be re-investigated before applying it to a different array configuration such as a different antenna spacing or frequency band.
For the comparison of experiments, more universal quantities, such as the electric field strength or the total radiation energy in a given frequency band \cite{Aab:2016eeq} should be used.

\section{Imaging of Air Showers}
\begin{figure}
  \centering
  \includegraphics[width=0.99\linewidth]{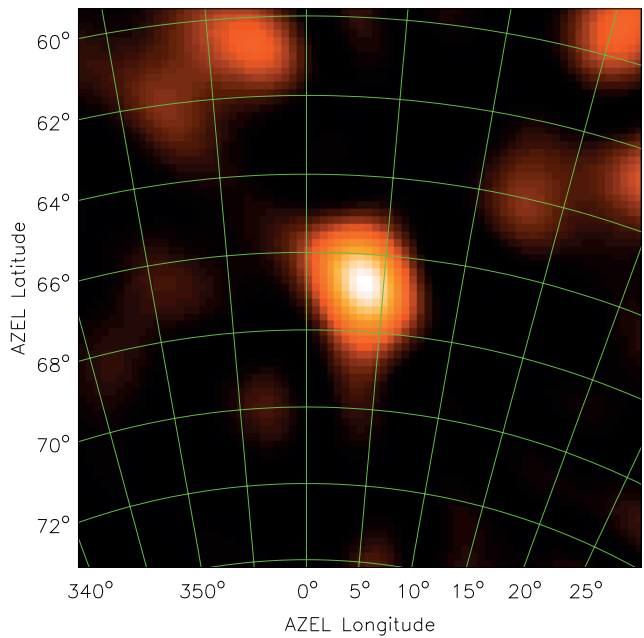}

  \caption{Image of the radio emission of an air-shower detected by LOPES (as originally published in \cite{FalckeNature2005}). 
  The bright spot in the center indicates the arrival direction of the air shower.}
  \label{fig_showerImage}
\end{figure}

LOPES provided the first successful radio images of air showers (Fig.\ref{fig_showerImage}, \cite{FalckeNature2005}). 
For a few $10\,$ns, the air shower is the brightest source in the sky.
The image clearly marks the arrival direction of the air showers.
The brightness is correlated to the size of the  electromagnetic shower component and, thus, to the energy of the primary particle. 
Moreover, the wavefront shape used to produce the image via beamforming contains information on the distance of the emission.
However, the radio images of showers were insufficient to reveal any of the substructure of the detected air showers.

Unlike images of extended sources in astronomy and unlike the air-shower images produced by the incoherent air-fluorescence technique, the radio images of the air showers observed by LOPES were just bright points with no sub-structure. 
Since the shower front and the radio signal both propagate with approximately the speed of light, the radio signal detected by LOPES is time compressed to a single pulse containing the emission from the whole shower development. 
This is also true for air-Cherenkov light emitted by air showers, but since the optical wavelengths are small, at least lateral substructure is visible in Cherenkov-light images of air showers. 
The radio emission, however, originates mostly from a region around the shower axis smaller than one wavelength. 
Therefore, it is not surprising that no substructures were visible in the radio images - at least at the LOPES frequency band of $40-80\,$MHz and at small viewing angles corresponding to the small axis distances where LOPES observed the showers (at larger viewing angles corresponding to larger axis distances it might be easier to resolve the structure of the shower, but the signal is significantly weaker).  
On top of these intrinsic difficulties in radio imaging of air showers, there are side lobes of the instrument and secondary maxima of the cross-correlation beam which lead to apparent structure in the image. 
These difficulties still need to be resolved before radio imaging can be applied to the open questions in air-shower and cosmic-ray physics. 

\section{Observation of Thunderstorms}
LOPES featured a special mode for the observation of thunderstorms making utilizing the long buffer time of several ms in the transient buffer boards used for data acquisition.
In addition to the radio signals of the air showers, that were altered by the atmospheric electric fields of the thunderclouds \cite{BuitinkLOPESthunderstorms2007}, also the direct radio emission of lightning was observed \cite{2011ApelLOPES_Thunderstorm}.
Due to the limited statistics and small size of the LOPES and KASCADE-Grande arrays, no conclusion could be made whether air-showers initiate lightning. 
Nonetheless, the idea of studying thunderclouds by the radio signals detected from air showers was developed further and successfully applied with higher accuracy at LOFAR \cite{SchellartLOFARthunderstorm2015}.
Recently the enhancement of radio signals of air-showers during thunderstorm conditions was also confirmed at lower frequencies of a few MHz \cite{Charrier:2019zey}.  

\section{Analysis Software}
To our knowledge, LOPES was the first radio array for air showers whose analysis software was made available with open access.
Some components of the software were shared between LOPES and LOFAR, which was easy because both software resides in the same repository. 
However, we do not know about any external users of the LOPES software. 
There are at least two reasons why our software was difficult to transfer to other experiments.
First, the LOPES analysis pipeline was designed specifically for the needs of digital radio interferometry, and other antenna arrays mostly use non-interferometric approaches. 
Second, and most importantly, the LOPES software was hard to maintain and hard to compile on new systems. 
The software makes use of several external libraries which were not well maintained and did not compile easily on more modern systems. 
The examples of the Offline software of the Pierre Auger Collaboration \cite{RadioOffline2011} and of the newer NuRadioReco software \cite{GlaserNuRadioRec2019} show that a different software design and strict coding rules greatly facilitate the use of the software for other experiments - to the benefit of the community. 
Consequently, we recommend that new software developed for future experiments should follow a modular approach and restrict the use of external libraries to those widely used and likely to be maintained for a long time.

\section{Frequency Band}
\label{sec_freqBand}
The frequency band of LOPES of $40-80\,$MHz was chosen for technical considerations and taking into account the limited knowledge of the radio emission of air showers that was available when LOPES was designed. 
It was not yet known that the coherence conditions lead to strong emission up to a few GHz at the Cherenkov angle \cite{CROME_PRL2014} despite the radio emission not being Cherenkov light \cite{James:2010vm}.
Since the Cherenkov ring has a diameter of about $100-150\,$m around the shower axis for the zenith angle range until $45^\circ$ and the altitude of $110\,$m of LOPES, a higher frequency band would likely have had improved the signal-to-noise ratio.

Meanwhile, the progress in high-fidelity simulations of the radio emission enabled dedicated design studies for future experiments. 
Therefore, we have learned that depending on the science case, the instrumented area and the antenna spacing, other frequency bands might be better.
In particular, higher frequency bands up to several $100\,$MHz can improve the signal-to-noise ratio and lower the detection threshold \cite{Balagopal2017}.
This is now taken into account in the design of future radio arrays \cite{SchroederICRC2019_SouthPoleRadio,GRAND_WhitePaper2020}.

\section{Practical Advice}
Last but not least, we learned several practical lessons in operating the experiment, which we list here in the hope that others may benefit from our experience:
\begin{itemize}
\item A graphical user interface helps to understand experimental issues quickly.
\item A monitoring website providing a quick overview of the experimental status is very helpful.
\item It occasionally happened that during maintenance or deployment the polarization channels of an antenna were accidentally swapped. Therefore, after each maintenance operation, this should be checked by appropriate monitoring tools, e.g., by the relative strengths of RFI or beacon lines in the frequency spectra recorded by the antenna channels.
\item Most difficult to detect were polarity flips of channels (corresponding to an accidental rotation of $180^\circ$ of the antenna). Although an antenna looks symmetrical, polarity flips can degrade the performance of the array, in particular when combining antennas in an interferometric analysis.
\item All cables should be deployed on or preferably under ground, but not in the air. Otherwise, electrical ground loops may impact the measurements.
\item RFI emitted by other electronics or detectors is difficult to mitigate by shielding alone because Faraday cages only help to a limited extent. For LOPES, it was critical that the distance between the antennas and the closest KASCADE particle detectors was larger than one wavelength. This enabled us to distinguish the radio signals emitted by air showers and the RFI 'signals' by the particle detectors by timing.
\end{itemize}

\bibliographystyle{hieeetr}
\bibliography{lopesFinalResults}   

\begin{thebibliography}{10}

\bibitem{jelley}
J.~V. {Jelley}, J.~H. {Fruin}, N.~A. {Porter}, {\em et~al.}, ``{Radio Pulses
  from Extensive Cosmic-Ray Air Showers},'' {\em Nature}, vol.~205,
  pp.~327--328, 1965.

\bibitem{Allan1971}
H.~{Allan}, ``{Radio Emission From Extensive Air Showers},'' {\em Progress in
  Elementary Particle and Cosmic Ray Physics}, vol.~10, pp.~171--302, 1971.

\bibitem{Vedeneev:2009zz}
O.~Vedeneev, ``{Depth of the maximum of extensive air showers and mass
  composition of primary cosmic radiation at an energy of 4 x 10**17 eV
  according to data on radioemission from extensive air showers},'' {\em Phys.
  Atom. Nucl.}, vol.~72, pp.~250--256, 2009.

\bibitem{HuegeReview2016}
{T.~Huege}, ``{Radio detection of cosmic ray air showers in the digital era},''
  {\em Physics Reports}, vol.~620, p.~1, 2016.

\bibitem{SchroederReview2016}
{F.G.~Schr\"oder}, ``{Radio detection of Cosmic-Ray Air Showers and High-Energy
  Neutrinos},'' {\em Prog. Part. Nucl. Phys.}, vol.~93, pp.~1--68, 2017.

\bibitem{FalckeNature2005}
{H.~Falcke, et al.~- LOPES Coll.}, ``{Detection and imaging of atmospheric
  radio flashes from cosmic ray air showers},'' {\em Nature}, vol.~435,
  pp.~313--316, 2005.

\bibitem{ArdouinBelletoileCharrier2005}
{D.~Ardouin, et al.~- CODALEMA Coll.}, ``{Radio-detection signature of
  high-energy cosmic rays by the CODALEMA experiment},'' {\em Nucl. Inst. Meth.
  A}, vol.~555, pp.~148--163, 2005.

\bibitem{HuegeCoREAS_ARENA2012}
T.~{Huege}, M.~{Ludwig}, and C.~{James}, ``{Simulating radio emission from air
  showers with CoREAS},'' {\em AIP Conf. Proceedings}, vol.~1535, pp.~128--132,
  2013.

\bibitem{Alvarez_ZHAires_2012}
J.~{Alvarez-Mu\~niz} {\em et~al.}, ``{Monte Carlo simulations of radio pulses
  in atmospheric showers using ZHAireS},'' {\em Astropart. Phys.}, vol.~35,
  pp.~325--341, 2012.

\bibitem{SELFAS2_ARENA2012}
{V.~Marin}, ``{SELFAS2 : radio emission from cosmic ray air showers. Effect of
  realistic air refractive index},'' {\em AIP Conf. Proceedings}, vol.~1535,
  p.~148, 2013.

\bibitem{Werner_EVA2012}
K.~{Werner} {\em et~al.}, ``{A Realistic Treatment of Geomagnetic Cherenkov
  Radiation from Cosmic Ray Air Showers},'' {\em Astropart. Phys.}, vol.~37,
  p.~5, 2012.

\bibitem{HuegeAERA_UHECR2018}
{T.~Huege for the Pierre Auger Coll.}, ``{Radio detection of cosmic rays with
  the Auger Engineering Radio Array},'' {\em EPJ Web Conf.}, vol.~210,
  p.~05011, 2019, 1905.04986.

\bibitem{SchellartLOFAR2013}
{P.~Schellart, et al.~- LOFAR Coll.}, ``{Detecting cosmic rays with the LOFAR
  radio telescope},'' {\em Astronomy \& Astrophysics}, vol.~560, p.~A98, 2013.

\bibitem{TREND2019}
D.~"Charrier and others", ``{Autonomous radio detection of air showers with the
  TREND50 antenna array},'' {\em Astropart. Phys.}, vol.~110, pp.~15--29, 2019,
  1810.03070.

\bibitem{TunkaRex_Xmax2016}
{P.A.~Bezyazeekov, et al.~- Tunka-Rex Coll.}, ``{Radio measurements of the
  energy and depth of maximum of cosmic-ray air showers by Tunka-Rex},'' {\em
  JCAP}, vol.~01, p.~052, 2016.

\bibitem{NiglDirection2008}
{A.~Nigl, et al.~- LOPES Coll.}, ``{Direction identification in radio images of
  cosmic-ray air showers detected with LOPES and KASCADE},'' {\em Astronomy \&
  Astrophysics}, vol.~487, pp.~781--788, 2008.

\bibitem{2014ApelLOPES_MassComposition}
{W.D.~Apel, et al.~- LOPES Coll.}, ``{Reconstruction of the energy and depth of
  maximum of cosmic-ray air-showers from LOPES radio measurements},'' {\em
  Phys. Rev. D}, vol.~90, p.~062001, 2014.

\bibitem{BuitinkLOFAR_Xmax2014}
{S.~Buitink et al.~-LOFAR Coll.}, ``{Method for high precision reconstruction
  of air shower $X_\mathrm{max}$ using two-dimensional radio intensity
  profiles},'' {\em Phys. Rev. D}, vol.~90, p.~082003, 2014.

\bibitem{AERA_XmaxMethods_ARENA2016}
{F. Gat\'e for the Pierre Auger Coll.}, ``{$X_\mathrm{max}$ reconstruction from
  amplitude information with AERA},'' {\em EPJ Web Conf.}, vol.~135, p.~01007,
  2017, 1609.06510.

\bibitem{TunkaRexXmaxPRD2018}
{P.A.~Bezyazeekov, et al.~- Tunka-Rex Coll.}, ``{Reconstruction of cosmic ray
  air showers with Tunka-Rex data using template fitting of radio pulses},''
  {\em Phys.~Rev.~D}, vol.~97, p.~122004, 2018.

\bibitem{2011ApelLOPES_Thunderstorm}
{W.D.~Apel, et al.~- LOPES Coll.}, ``{Thunderstorm observations by air-shower
  radio antenna arrays},'' {\em Advances in Space Research}, vol.~48,
  pp.~1295--1303, 2011.

\bibitem{NellesLOFAR_measuredLDF2015}
{A.~Nelles, et al.~- LOFAR Coll.}, ``{The radio emission pattern of air showers
  as measured with LOFAR - a tool for the reconstruction of the energy and the
  shower maximum},'' {\em JCAP}, vol.~05, p.~018, 2015.

\bibitem{AugerAERAenergy2015}
{Pierre Auger Coll.}, ``{Energy Estimation of Cosmic Rays with the Engineering
  Radio Array of the Pierre Auger Observatory},'' {\em Phys. Rev. D}, vol.~93,
  p.~122005, 2016.

\bibitem{KostuninTheory2015}
D.~{Kostunin} {\em et~al.}, ``{Reconstruction of air-shower parameters for
  large-scale radio detectors using the lateral distribution},'' {\em
  Astropart. Phys.}, vol.~74, p.~79, 2016.

\bibitem{KahnLerche1966}
F.~D. {Kahn} and I.~{Lerche}, ``{Radiation from cosmic ray air showers},'' in
  {\em Proceedings of the Royal Society of London. Series A, Mathematical and
  Phys. Sciences}, vol.~289, p.~206, 1966.

\bibitem{FalckeGorham2003}
H.~{Falcke} and P.~W. {Gorham}, ``{Detecting radio emission from cosmic ray air
  showers and neutrinos with a digital radio telescope},'' {\em Astropart.
  Phys.}, vol.~19, pp.~477--494, 2003.

\bibitem{Askaryan1962}
G.~A. {Askaryan}, ``{Excess negative charge of an electron-photon shower and
  its coherent radio emission},'' {\em Soviet Physics JETP}, vol.~14, p.~441,
  1962.

\bibitem{AugerAERApolarization2014}
{Pierre Auger Coll.}, ``{Probing the radio emission from cosmic-ray-induced air
  showers by polarization measurements},'' {\em Phys. Rev. D}, vol.~89,
  p.~052002, 2014.

\bibitem{CODALEMAchargeExcess2015}
{A.~Bell\'{e}toile, et al.~- CODALEMA Coll.}, ``{Evidence for the charge-excess
  contribution in air shower radio emission observed by the CODALEMA
  experiment},'' {\em Astropart. Phys.}, vol.~69, pp.~50--60, 2015.

\bibitem{SchellartLOFARpolarization2014}
{P.~Schellart, et al.~- LOFAR Coll.}, ``{Polarized radio emission from
  extensive air showers measured with LOFAR},'' {\em JCAP}, vol.~10, p.~014,
  2014.

\bibitem{LinkPhDThesis2016}
K.~{Link}, ``{Improved analysis of air shower radio emission with the LOPES
  experiment},'' {PhD Thesis}, Karlsruhe Inst.~Tech., Germany, 2016.
\newblock {doi:10.5445/IR/1000062597}.

\bibitem{LOPES_ICRC2015_Link}
{K.Link for the LOPES Coll.}, ``{Revised absolute amplitude calibration of the
  LOPES experiment},'' {\em Proceedings of Science}, vol.~236, p.~311, 2016.
\newblock PoS(ICRC2015)311.

\bibitem{LOPES_ICRC2015_Schroeder}
{F.G.~Schr\"oder for the LOPES Coll.}, ``{New results of the digital radio
  interferometer LOPES},'' {\em Proceedings of Science}, vol.~236, p.~317,
  2016.
\newblock PoS(ICRC2015)317.

\bibitem{LOPES_ICRC2017}
{F.G.~Schr\"oder for the LOPES Coll.}, ``{Interferometric Radio Measurements of
  Air Showers with LOPES: Final Results},'' {\em Proceedings of Science},
  vol.~301, p.~458, 2018.
\newblock PoS(ICRC2017)458.

\bibitem{IsarARENA_2008}
{G.~Isar for the LOPES Coll.}, ``{Radio emission of energetic cosmic ray air
  showers: Polarization measurements with LOPES},'' {\em Nucl. Inst. Meth. A},
  vol.~604, pp.~S81--S84, 2009.

\bibitem{HuberPhDThesis2014}
D.~{Huber}, ``{Analysing the electric field vector of air shower radio
  emission},'' {PhD Thesis}, Karlsruhe Inst.~Tech., Germany, 2014.
\newblock {doi:10.5445/IR/1000043289}.

\bibitem{PetrovicLOPESinclined2007}
{J.~Petrovic, et al.~- LOPES Coll.}, ``{Radio emission of highly inclined
  cosmic ray air showers measured with LOPES},'' {\em Astronomy \&
  Astrophysics}, vol.~462, pp.~389--395, 2007.

\bibitem{2013ApelLOPESlateralComparison}
{W.D.~Apel, et al.~- LOPES Coll.}, ``{Comparing LOPES measurements of
  air-shower radio emission with REAS 3.11 and CoREAS simulations},'' {\em
  Astropart. Phys.}, vol.~50-52, pp.~76--91, 2013.

\bibitem{ANITA_CR_PRL_2010}
{S. Hoover, et al.~- ANITA Coll.}, ``{Observation of Ultrahigh-Energy Cosmic
  Rays with the ANITA Balloon-Borne Radio Interferometer},'' {\em Phys. Rev.
  Lett.}, vol.~105, p.~151101, 2010.

\bibitem{AERAinclined2018}
{The Pierre Auger Collaboration}, ``{Observation of inclined EeV air showers
  with the radio detector of the Pierre Auger Observatory},'' {\em JCAP},
  vol.~10, p.~026, 2018, 1806.05386.

\bibitem{BuitinkLOPESthunderstorms2007}
{S.~Buitink et al.~-LOPES Coll.}, ``{Amplified radio emission from cosmic ray
  air showers in thunderstorms},'' {\em Astronomy \& Astrophysics}, vol.~467,
  p.~385, 2007.

\bibitem{Apel:2006pv_LOPESdistantEvents}
{W.D.~Apel, et al.~- LOPES Coll.}, ``{Progress in Air Shower Radio
  Measurements: Detection of Distant Events},'' {\em Astropart. Phys.},
  vol.~26, pp.~332--340, 2006, astro-ph/0607495.

\bibitem{NehlsHakenjosArts2007}
S.~{Nehls} {\em et~al.}, ``{Amplitude calibration of a digital radio antenna
  array for measuring cosmic ray air showers},'' {\em {Nucl. Inst. Meth. A}},
  vol.~589, pp.~350 -- 361, 2008.

\bibitem{2015ApelLOPES_improvedCalibration}
{W.D.~Apel, et al.~- LOPES Coll.}, ``{Improved absolute calibration of LOPES
  measurements and its impact on the comparison with REAS 3.11 and CoREAS
  simulations},'' {\em Astropart. Phys.}, vol.~75, p.~72, 2016.

\bibitem{SchroederTimeCalibration2010}
{F.G.~Schr\"oder} {\em et~al.}, ``{New method for the time calibration of an
  interferometric radio antenna array},'' {\em Nucl. Inst. Meth. A}, vol.~615,
  pp.~277--284, 2010.

\bibitem{Ludwig:2010pf}
M.~Ludwig and T.~Huege, ``{REAS3: Monte Carlo simulations of radio emission
  from cosmic ray air showers using an 'end-point' formalism},'' {\em
  Astropart. Phys.}, vol.~34, pp.~438--446, 2011, 1010.5343.

\bibitem{2010ApelLOPESlateral}
{W.D.~Apel, et al.~- LOPES Coll.}, ``{Lateral distribution of the radio signal
  in extensive air showers measured with LOPES},'' {\em Astropart. Phys.},
  vol.~32, pp.~294--303, 2010.

\bibitem{2014ApelLOPES_wavefront}
{W.D.~Apel, et al.~- LOPES Coll.}, ``{The wavefront of the radio signal emitted
  by cosmic ray air showers},'' {\em JCAP}, vol.~09, p.~025, 2014.

\bibitem{CorstanjeLOFAR_wavefront2014}
{A.~Corstanje et al.~-LOFAR Coll.}, ``{The shape of the radio wavefront of
  extensive air showers as measured with LOFAR},'' {\em Astropart. Phys.},
  vol.~61, pp.~22--31, 2015.

\bibitem{NiglFrequencySpectrum2008}
{A.~Nigl, et al.~- LOPES Coll.}, ``{Frequency spectra of cosmic ray air shower
  radio emission measured with LOPES},'' {\em Astronomy \& Astrophysics},
  vol.~488, pp.~807--817, 2008.

\bibitem{2012ApelLOPES_MTD}
{W.D.~Apel, et al.~- LOPES Coll.}, ``{Experimental evidence for the sensitivity
  of the air-shower radio signal to the longitudinal shower development},''
  {\em Phys. Rev. D}, vol.~85, p.~071101(R), 2012.

\bibitem{AntoniApelBadea2003}
{T.~Antoni, et al.~- KASCADE Coll.}, ``{The Cosmic-Ray Experiment KASCADE},''
  {\em Nucl. Inst. Meth. A}, vol.~513, pp.~490--510, 2003.

\bibitem{Apel2010KASCADEGrande}
{W.D.~Apel, et al.~- KASCADE-Grande Coll.}, ``{The KASCADE-Grande
  experiment},'' {\em Nucl. Inst. Meth. A}, vol.~620, pp.~202--216, 2010.

\bibitem{ScholtenLOFAR_circPol2016}
{O.~Scholten et al.~-LOFAR Coll.}, ``{Measurement of the circular polarization
  in radio emission from extensive air showers confirms emission mechanisms},''
  {\em Phys. Rev. D}, vol.~94, p.~103010, 2016.

\bibitem{2012ApelLOPES3D}
{W.D.~Apel, et al.~- LOPES Coll.}, ``{LOPES-3D: An antenna array for full
  signal detection of air-shower radio emission},'' {\em Nucl. Inst. Meth. A},
  vol.~696, pp.~100--109, 2012.

\bibitem{HornefferThesis2006}
A.~{Horneffer}, ``{Measuring Radio Emission from Cosmic Ray Air Showers with a
  Digital Radio Telescope},'' {PhD Thesis}, Rheinische
  Friedrich-Wilhelms-Universit\"at Bonn, Germany, 2006.
\newblock {http://nbn-resolving.de/urn:nbn:de:hbz:5N-07819}.

\bibitem{CROME_PRL2014}
{R.~Smida, et al.~- CROME Coll.}, ``{First Experimental Characterization of
  Microwave Emission from Cosmic Ray Air Showers},'' {\em Phys. Rev. Lett.},
  vol.~113, p.~221101, 2013.

\bibitem{LOPESsoftware2017}
{LOPES Collaboration}, ``{LOPES analysis software (includes calibration
  tables)},'' {Open-source software written in C++}, 2017.
\newblock {http://usg.lofar.org/svn/code/branches/cr-tools-stable}.

\bibitem{SchroederThesis2011}
F.~G. {Schr\"oder}, ``{Instruments and Methods for the Radio Detection of High
  Energy Cosmic Rays},'' {PhD Thesis; published by Springer Theses}, Karlsruhe
  Inst.~Tech., Germany, 2011.
\newblock {doi:10.5445/IR/1000022313 and doi:10.1007/978-3-642-33660-7}.

\bibitem{HuegeARENA_LOPESSummary2010}
{T.~Huege, et al.~- LOPES Coll.}, ``{The LOPES experiment - recent results,
  status and perspectives},'' in {\em Nucl. Inst. Meth. A; Proceedings of the
  ARENA 2010 conference, Nantes, France}, vol.~662, Supplement 1, pp.~S72--S79,
  2012.

\bibitem{Schoorlemmer:2020low}
H.~Schoorlemmer and W.~R. Carvalho, ``{Radio interferometry applied to the
  observation of cosmic-ray induced extensive air showers},'' 2020, 2006.10348.

\bibitem{Horneffer:2008hma}
{A.~Horneffer for the LOPES Coll.}, ``{Detecting radio pulses from air
  showers},'' in {\em {2008 IEEE Nuclear Science Symposium and Medical Imaging
  Conference and 16th International Workshop on Room-Temperature Semiconductor
  X-Ray and Gamma-Ray Detectors}}, pp.~3339--3346, 2008.

\bibitem{SchroederNoise2010}
{F.G.~Schr\"oder, et al.~- LOPES Coll.}, ``{On noise treatment in radio
  measurements of cosmic ray air showers},'' {\em Nucl. Inst. Meth. A},
  vol.~662, Supplement 1, pp.~S238--S241, 2012.

\bibitem{Huege:2008tn}
T.~Huege, R.~Ulrich, and R.~Engel, ``{Energy and composition sensitivity of
  geosynchrotron radio emission from cosmic ray air showers},'' {\em Astropart.
  Phys.}, vol.~30, p.~96, 2008, 0806.1161.

\bibitem{KCDC2018}
{A. Haungs, et~al. - KASCADE-Grande Coll.}, ``{The KASCADE Cosmic-ray Data
  Centre KCDC: Granting Open Access to Astroparticle Physics Research Data},''
  {\em Eur. Phys. J. C}, vol.~78, no.~9, p.~741, 2018, 1806.05493.

\bibitem{HuegeSKA_ICRC2015}
T.~{Huege} {\em et~al.}, ``{High-precision measurements of extensive air
  showers with the SKA},'' {\em Proceedings of Science}, vol.~236, p.~309,
  2016.
\newblock PoS(ICRC2015)309.

\bibitem{LenokICRC2019}
{V.~Lenok for the Tunka-Rex Coll.}, ``{Modeling the Aperture of Radio
  Instruments for Air-Shower Detection},'' {\em PoS}, vol.~331, p.~418, 2020,
  1909.01945.

\bibitem{LOFARNature2016}
{S.~Buitink, et al.~- LOFAR Coll.}, ``{A large light-mass component of cosmic
  rays at $10^{17}-10^{17.5}$ electronvolts from radio observations},'' {\em
  Nature}, vol.~531, p.~70, 2016.

\bibitem{SchroederTunkaRex_PISA2015}
{F.G.~Schr\"oder, et al.~Tunka-Rex Coll.}, ``{The Tunka radio extension
  (Tunka-Rex): Radio measurements of cosmic rays in Siberia},'' {\em Nucl.
  Inst. Meth. A}, vol.~824, p.~652, 2016.

\bibitem{AERAairplanePaper2015}
{Pierre Auger Coll. et al.}, ``{Nanosecond-level time synchronization of
  autonomous radio detector stations for extensive air showers},'' {\em JINST},
  vol.~11, p.~P01018, 2016.

\bibitem{TunkaRex_NIM_2015}
{P.A.~Bezyazeekov, et al.~- Tunka-Rex Coll.}, ``{Measurement of cosmic-ray air
  showers with the Tunka Radio Extension (Tunka-Rex)},'' {\em Nucl. Inst. Meth.
  A}, vol.~802, pp.~89--96, 2015.

\bibitem{AERAantennaCalibration2017}
{Pierre Auger Coll. et al.}, ``{Calibration of the Logarithmic-Periodic Dipole
  Antenna (LPDA) Radio Stations at the Pierre Auger Observatory using an
  Octocopter},'' {\em JINST}, vol.~12, p.~T10005, 2017.

\bibitem{NellesLOFAR_calibration2015}
{A.~Nelles et al.}, ``{Calibrating the absolute amplitude scale for air showers
  measured at LOFAR},'' {\em JINST}, vol.~10, p.~P11005, 2015.

\bibitem{TunkaRexScale2016}
{W.D.~Apel, et al.~- LOPES and Tunka-Rex Colls.}, ``{A comparison of the
  cosmic-ray energy scales of Tunka-133 and KASCADE-Grande via their radio
  extensions Tunka-Rex and LOPES},'' {\em Physics Lett. B}, vol.~763,
  pp.~179--185, 2016.

\bibitem{GRAND_WhitePaper2020}
{J.~Alvarez-Mu\~niz, et al. - GRAND Coll.}, ``{The Giant Radio Array for
  Neutrino Detection (GRAND): Science and Design},'' {\em Sci. China Phys.
  Mech. Astron.}, vol.~63, no.~1, p.~219501, 2020, 1810.09994.

\bibitem{SchroderAstro2020}
F.~G. {Schr\"oder} {\em et~al.}, ``{High-Energy Galactic Cosmic Rays (Astro2020
  Science White Paper)},'' {\em Bull. Am. Astron. Soc.}, vol.~51, p.~131, 3
  2019, 1903.07713.

\bibitem{SarazinAstro2020}
F.~Sarazin {\em et~al.}, ``{What is the nature and origin of the highest-energy
  particles in the universe?},'' {\em Bull. Am. Astron. Soc.}, vol.~51, no.~3,
  p.~93, 2019, 1903.04063.

\bibitem{HoltEPJ_C}
E.~M. Holt {\em et~al.}, ``{Enhancing the cosmic-ray mass sensitivity of
  air-shower arrays by combining radio and muon detectors},'' {\em Eur. Phys.
  J. C}, vol.~79, no.~5, p.~371, 2019, 1905.01409.

\bibitem{HaungsIceCubeSurfaceUHECR2018}
{A.~Haungs for the IceCube Coll.}, ``{A Scintillator and Radio Enhancement of
  the IceCube Surface Detector Array},'' {\em EPJ Web Conf.}, vol.~210,
  p.~06009, 2019, 1903.04117.

\bibitem{SchroederICRC2019_SouthPoleRadio}
{F.G.~Schr\"oder for the IceCube Coll.}, ``{Science Case of a Scintillator and
  Radio Surface Array at IceCube},'' {\em PoS}, vol.~358, p.~418, 2020,
  1908.11469.

\bibitem{HorandelAugerRadioUpgradeUHECR2018}
{J.~H\"orandel for the Pierre Auger Coll.}, ``{Precision measurements of cosmic
  rays up to the highest energies with a large radio array at the Pierre Auger
  Observatory},'' {\em EPJ Web Conf.}, vol.~210, p.~06005, 2019.

\bibitem{PontICRC2019_AugerRadioUpgrade}
{B.~Pont for the Pierre Auger Coll.}, ``{A Large Radio Detector at the Pierre
  Auger Observatory - Measuring the Properties of Cosmic Rays up to the Highest
  Energies},'' {\em PoS}, vol.~358, p.~395, 2020, 1909.09073.

\bibitem{SchmidtIcrc2009}
{A.~Schmidt, et al.~- LOPES Coll.}, ``{Self-Trigger for Radio Detection of
  UHCR},'' in {\em {Proceedings of the 31st ICRC, {\L}\'{o}d\'{z}, Poland}},
  no.~1124, 2009.
\newblock {FZKA report 7516,
  http://icrc2009.uni.lodz.pl/proc/pdf/icrc1124.pdf}.

\bibitem{RAugerSelfTrigger2012}
{Pierre Auger Coll., et al.}, ``{Results of a self-triggered prototype system
  for radio-detection of extensive air showers at the Pierre Auger
  Observatory},'' {\em JINST}, vol.~7, p.~P11023, 2012.

\bibitem{Kelley:2013rma}
{J.L.~Kelley for the Pierre Auger Coll.}, ``{Data Acquisition, Triggering, and
  Filtering at the Auger Engineering Radio Array},'' {\em Nucl. Instrum. Meth.
  A}, vol.~725, pp.~133--136, 2013, 1205.2104.

\bibitem{Revenu:2019imx}
{B.~Revenu for the CODALEMA Coll.}, ``{Current status of the CODALEMA/EXTASIS
  experiments},'' {\em J. Phys. Conf. Ser.}, vol.~1181, no.~1, p.~012029, 2019.

\bibitem{ARIANNA_2017}
{S.W. Barwick, et~al. - ARIANNA Coll.}, ``{Radio detection of air showers with
  the ARIANNA experiment on the Ross Ice Shelf},'' {\em Astropart. Phys.},
  vol.~90, pp.~50--68, 2017, 1612.04473.

\bibitem{LinkDiplomaThesis2006}
K.~{Link}, ``{Measurement of kHz Radio Emission of Air Showers},'' {Diploma
  Thesis}, Karlsruhe Institute of Technology (KIT), Germany, 2009.
\newblock {https://www.astro.ru.nl/lopes/publications/phd\_theses}.

\bibitem{Huege:2019ufo}
{T.~Huege and C.B.~Welling for the Pierre Auger Coll.}, ``{Reconstruction of
  air-shower measurements with AERA in the presence of pulsed radio-frequency
  interference},'' {\em EPJ Web Conf.}, vol.~216, p.~03007, 2019, 1906.05148.

\bibitem{2005AIPC..745..770G}
A.~W. {Guthmann} and B.~{Thid{\'e}}, ``{The LOIS project and astrophysics},''
  in {\em High Energy Gamma-Ray Astronomy} (F.~A. {Aharonian}, H.~J.
  {V{\"o}lk}, and D.~{Horns}, eds.), vol.~745 of {\em American Institute of
  Physics Conference Series}, pp.~770--773, Feb. 2005.

\bibitem{Revenu:2017rui}
{B.~Revenut, et al.~- CODALEMA Coll.}, ``{The CODALEMA/EXTASIS experiment: a
  multi-scale and multi-wavelength instrument for radio-detection of extensive
  air-showers},'' {\em PoS}, vol.~ICRC2017, p.~416, 2018.

\bibitem{Aab:2016eeq}
A.~Aab {\em et~al.}, ``{Measurement of the Radiation Energy in the Radio Signal
  of Extensive Air Showers as a Universal Estimator of Cosmic-Ray Energy},''
  {\em Phys. Rev. Lett.}, vol.~116, no.~24, p.~241101, 2016, 1605.02564.

\bibitem{SchellartLOFARthunderstorm2015}
{P.~Schellart, et al.~- LOFAR Coll.}, ``{Probing Atmospheric Electric Fields in
  Thunderstorms through Radio Emission from Cosmic-Ray-Induced Air Showers},''
  {\em Phys. Rev. Lett.}, vol.~114, p.~165001, 2015.

\bibitem{Charrier:2019zey}
{D.~Charrier, et al.~- CODALEMA Coll.}, ``{Radio detection of cosmic rays in
  [1.7-3.7] MHz: The EXTASIS experiment},'' {\em Astropart. Phys.}, vol.~113,
  pp.~6--21, 2019, 1903.02792.

\bibitem{RadioOffline2011}
{Pierre Auger Coll.}, ``{Advanced functionality for radio analysis in the
  Offline software framework of the Pierre Auger Observatory},'' {\em Nucl.
  Inst. Meth. A}, vol.~635, pp.~92--102, 2011.

\bibitem{GlaserNuRadioRec2019}
C.~Glaser {\em et~al.}, ``{NuRadioReco: A reconstruction framework for radio
  neutrino detectors},'' {\em Eur. Phys. J. C}, vol.~79, no.~6, p.~464, 2019,
  1903.07023.

\bibitem{James:2010vm}
C.~W. James, H.~Falcke, T.~Huege, and M.~Ludwig, ``{General description of
  electromagnetic radiation processes based on instantaneous charge
  acceleration in `endpoints'},'' {\em Phys. Rev. E}, vol.~84, p.~056602, 2011,
  1007.4146.

\bibitem{Balagopal2017}
A.~Balagopal~V., A.~Haungs, T.~Huege, and F.~G. Schroeder, ``{Search for
  PeVatrons at the Galactic Center using a radio air-shower array at the South
  Pole},'' {\em Eur. Phys. J.}, vol.~C78, no.~2, p.~111, 2018, 1712.09042.
\newblock [erratum: Eur. Phys. J.C78,no.12,1017(2018)].

\end{thebibliography}


\end{document}